\documentstyle{aipproc}
\begin{document}
\title{Towards a Loop Representation of  \\  Connection Theories  
defined  over
\\ a Super Lie Algebra}

\author{Luis F. Urrutia }
\address{Departamento de F\'\i sica \\
Universidad Aut\'onoma Metropolitana-Iztapalapa\\
Apartado Postal 55-534, 09340 M\'exico, D.F.\\
and \\
Instituto de Ciencias Nucleares\\
Universidad Nacional Aut\'onoma de M\'exico \\
Apartado Postal 70-543, 04510 M\'exico, D.F. }

\maketitle

\begin{abstract}
The purpose of this  contribution  is to review some aspects of   
the loop space
formulation of  pure gauge theories having  the connection defined  
over a Lie
algebra. The emphasis is focused on the discussion of the Mandelstam
identities, which provide the basic constraints upon  both  the  
classical and
the quantum  degrees of freedom  of  the theory.  In the case where the
connection is extended to be valued on a  super Lie algebra,  some  
new results
are presented
which  can be considered as first steps towards the construction of the
Mandelstam identities in this situation, which encompasses  such  
interesting
cases as supergravity in $3+1$ dimensions
together with $2+1$ super Chern-Simons theories, for example. Also,  
these ideas
 could be  useful in
the loop space formulation of fully supersymmetric theories.
\end{abstract}

\section*{1.  Introduction}

Gauge theories provide a successful  framework to unify the strong,  
weak and
electromagnetic
interactions in nature. The recent introduction of the Ashtekar  
variables in
general relativity
has permitted to reformulate the gravitational interaction in the  
form of a
standard  gauge theory, plus additional  constraints.  Gauge  
theories  where no
matter is included,  are
fully described in terms of
a single field:  the  connection $A_\mu$, which is the non-abelian
generalization of the standard electromagnetic potentials.  In  
general, gauge
theories are characterized  by  being invariant under  
transformations generated
by local  symmetries, i.e. transformations of the connection  which  
may be
differently chosen at each  point in space time.  This freedom has the
consequence of introducing arbitrary functions in the dynamics of the
connection.  In  this  way, only those specific fields constructed  
from the
connection  which are independent of these arbitrary functions,   
will be of
physical relevance: they  are called gauge invariant objects. From  
this point
of view one could say  that taking $A_\mu$ as the basic field
in a gauge theory is an unnecessary
complication. This observation has prompted many proposals to   
describe gauge
theories
starting {\it ab initio}  with only gauge invariant objects.

In this contribution  we will consider  the
so called loop representation approach, to be described  with more  
detail in
the sequel. This method has been very successful in the description of
non-perturbative aspects of
gauge theories and also  in providing  spectacular  advancements in the
problem of quantizing Einstein gravity. The basic underlying idea  
of the loop
representation
is that the Hilbert space of gauge invariant states can be spanned  
by states
which are labeled by loops.  A main feature of  the loop approach is the
redundancy
of the gauge invariant degrees of freedom: they constitute  an  
overcomplete set
which is
restricted by the so called Mandelstam identities.  These  
identities must be
taken into account,
either by imposing them over the physical states in a manner  
similar to  the
standard Dirac
constraints, or  by explicitly  solving them  in order to identify the
corresponding reduced phase space.
Even the preliminary problem of identifying the  complete set of  
Mandelstam
identities for a given
physical situation is not completely solved in the general case for gauge
theories over a Lie algebra.

In this work  we will  discuss  only theories  described by a connection
field, i.e.  those where matter fields are absent. These include   
pure Yang
Mills theories, pure  $2+1$ dimensional
Chern Simons theories and also gravity and supergravity in the Ashtekar
formulation, for example.  From the Hamiltonian point of view,  
these theories
are basically on the same footing. Each one is characterized by a  
specific
canonical Hamiltonian together with a given set of first-class  
constraints.
We
will be mostly concerned with the elucidation of the problem of the
Mandelstam identities, both in the standard  case and also  when
the connection of the  gauge theory  is defined over a super  Lie  
algebra.
Besides its
own interest, the latter situation is also relevant  as a first  
step in the
application of loop representation methods to fully supersymmetric gauge
theories, which involve matter fields as well. The discussion will   
be focused
at the classical level, without  considering in detail  the construction
of quantum states as functionals of loops.  These notes do not have the
pretense whatsoever  of being a review on the subject
and from the very beginning we apologize to those  authors whose  
work  we have
involuntarily
not cited.

The general organization of this paper is as follows: in Section 2   
we provide
a brief review
of the  standard formulation of  Yang Mills theories  in terms of the
connection. We recall the
Lagrangian formulation as well as the Hamiltonian approach, in  
which the loop
representation
is based.  Section 3 contains  the basic material necessary to extend the
connection to a  super  Lie algebra object and includes  the  
definition of a
supermatrix  together with those of the fundamental invariants  
under similarity
transformations: the supertrace and the superdeterminant.  An  
introduction to
the loop space  formulation of gauge theories is
presented in Section 4, where the basic concepts are reviewed to  
end up with
the
introduction of the gauge invariant degrees of freedom: the Wilson  
loops. Also,
the  construction
of the wave functionals  of the system in the loop space is briefly  
sketched.
Section 5 is devoted to the discussion of the overcompleteness of  
the loop
space variables and the
Mandelstam identities  are introduced to account for this  
redundancy, both in
the classical
and quantum case. The so called  generic Mandelstam identities are  
subsequently
derived
from the  identities arising from the application of the Cayley-Hamilton
theorem in  the specific representation where the connection lives.  
 Section 6
deals with the
extension of the previous ideas to the case where the connection is  
defined
over  a graded
Lie algebra. To begin with, we present an extension
of the Cayley-Hamilton theorem to the case of supermatrices.  
Besides its own
interest, our
motivation to deal with this mathematical problem is the claim  that the
algorithm
developed for the standard case in Section 5 can be extended  to this
situation, thus
providing the  corresponding  Mandelstam identities. The extension of the
Cayley-Hamilton theorem proceeds in two steps: (i) the identification and
definition of a  characteristic polynomial for supermatrices and  
(ii) the proof
that each supermatrix annihilates the null
polynomial previously defined.  Also, examples of  characteristic  
polynomials
for
supermatrices  are
given in  some simple cases.  Finally,  examples of  Mandelstam  
identities
constructed for some simple supermatrices, according to the
previous ideas, are presented.  We close with Section 7, which  
contains some
conclusions  and a brief
outlook of  some remaining open problems related to the topics  
discussed  in
this paper.

\section*{2. The connection approach to Yang Mills theories}

\subsection*{2.1. The Lagrangian Formulation}

Let us consider the Yang Mills theory corresponding to the compact   
group $G$,
which
 is characterized  by its Lie algebra $\cal A$ having the antihermitian 
generators $T^A, A=1,\dots,n$. They
satisfy  the
commutation relations $ \left[T^A, \ T^B \right]= f^{AB}{}_C \ T^C$,
where $f^{AB}{}_C$ are the structure constants of the group.
A standard representation of these generators is in terms  
of  ordinary matrices, with  additional  restrictions appropriate  
to the group
under consideration.

The basic object in this formulation  is the connection $A_\mu(x)$, 
which is a covariant vector field that lives in the Lie algebra of
$G$, i.e. $A_\mu(x)=A_{\mu C}(x)T^C $, where $\mu=0,1,2,3$ is a world
subindex. The infinitesimal local transformations 
$\delta \Theta(x)$ generated by  
the group can be written as
$\delta \Theta(x)=\delta \Theta(x)_C \ T^C $, where $\delta  
\Theta(x)_C $
are arbitrary numerical functions. Under such rotation, the
connection transforms as 
\begin{equation}
{\delta}_{\Theta} A_\mu(x)= \partial_\mu \delta \Theta(x) +
\left[A_\mu(x), \ \delta \Theta(x)   \right] := {\cal D}_\mu \delta  
\Theta(x) ,
\label{DELCON}
\end{equation}
which generalizes the gauge transformation of electrodynamics.
Here we have introduced also the covariant 
derivative of any object $M_C$ with one subindex in the Lie 
algebra, in the form
\begin{equation}
{\cal D}_\mu  M={\partial}_\mu \ M + 
\left[A_\mu(x), \  M   \right], \label{COVDER}
\end{equation}
where $M= M_C \ T^C $. In components, the above expression reads
\begin{equation}
\left( {\cal D}_\mu  M \right)_C=\partial_{\mu} M_C + f^{BD}{}_C   
A_{\mu B}
M_D.
\label{CDERCOM}
\end{equation}

The object  which is  covariant under  the local group of  
transformations  is
the field
strength $F_{\mu\nu}=F_{{\mu\nu}C} \ T^C$ defined by 
\begin{equation}
F_{\mu\nu}=\partial_\mu A_\nu-\partial_\nu A_\mu +  
\left[A_\mu, \ A_\nu \right], \label{FMUNU}
\end{equation}
which generalizes the electromagnetic field of electrodynamics. 
Under gauge rotations, $F_{\mu\nu}$ transforms 
covariantly , i.e. 
${\delta}_{\Theta} F_{\mu\nu}=\left[F_{\mu\nu}, \delta \Theta \right]$.
Besides, the field strength satisfies the Bianchi identity
\begin{equation}
{\cal D}_\mu F_{\nu\rho} + {\cal D}_\nu F_{\rho\mu} + 
{\cal D}_\rho F_{\mu\nu} = 0. \label{BID}
\end{equation}
Let us emphasize the basic property that the covariant derivative
(\ref{COVDER}) of any covariant object is itself a covariant object.

The dynamics of  the pure Yang Mills is described by the action
\begin{equation}
S= \int  d^4 x \  {\cal L} = \int d^4 x \ {1\over 4} Tr
(F_{\mu\nu}F^{\mu\nu}), 
\label{COVACT}
\end{equation}
where ${\cal L}$  is the Lagrangian density and
$ Tr( T^A \ T^B)= h^{AB}$ is an invariant 
symmetrical tensor in the corresponding representation of the Lie  
algebra.
For a semisimple group (i.e. one having no $U(1)$ invariant  
subgroups) it is
possible
to choose $h^{AB}=- \delta^{AB}$.  
The world indices are raised and lowered by the
Minskowsky metric $\eta_{\mu\nu}= diag(1,-1,-1,-1,-1)$.
The action (\ref{COVACT})
is invariant under the gauge transformations (\ref{DELCON}). Finally,
the resulting equations of motion are
\begin{equation}
{\cal D}_\mu \ F^{\mu\nu}=0. \label{EQMO}
\end{equation}

\subsection*{2.2 The Hamiltonian Formulation}

This formulation corresponds to the  $ (1+3)$ splitting of  Minkowsky
space-time, where we
select three-dimensional hypersurfaces of constant time. It is  
convenient to
introduce the chromoelectric, $E_i$, and chromomagnetic, $B^i$,  
fields in the
following way
\begin{equation}
E_i=F_{0i}= {\dot A}_i-{\cal D}_i A_0, \quad B^i={1\over2}\epsilon^{ijk}
F_{jk}.
\label{FIELDS}
\end{equation}
Here, $i=1,2,3 $ labels the components of a vector living on the
three-dimensional
hypersurface and the dot denotes the time derivative evaluated on that
hypersurface.
The covariant derivative ${\cal D}_i$ corresponds to the definition
(\ref{COVDER})
restricted again to the hypersurface,  through the conecction $A_i$.
Introducing the canonical momenta
\begin{equation}
\Pi^{i}{}_C:=
{ \partial {\cal L}\over \partial {\dot
A}_{i C}}=- E^i{}_C=E_{iC}
\label{MOM}
\end{equation}
and defining the Hamiltonian density
${\cal H}=\Pi^i{}_C  {\dot A}_{iC}- {\cal L}$,  we obtain the action
\begin{equation}
S= \int dt \int d^3x \  \left( {\Pi}^i{}_C  {\dot A}_{iC} - 
{1\over 2} ( E^2 + B^2) +
A_{0C} ({\cal D}_i E^i)_C \right), \label{HAMAC}
\end{equation}
where $ E^2=E^i{}_CE^i{}_C \geq 0 $ and analogously for $B^2$.
The above expression  arises after an additional  integration by  
parts, which
can be
performed in virtue of the
following properties of the covariant derivative: (i) \ $ {\cal  
D}_i (MN)= 
({\cal
D}_i M)N +
M ({\cal D}_i N)$ and (ii) \  $Tr( {\cal D}_i M)=\partial_i Tr(M)$.
The action (\ref{HAMAC}) tells us that the phase space variables
are the coordinates $A_{iC}({\vec x},t) $, together with their 
conjugated momenta
$\Pi^j{}_D({\vec y},t) =E_{jD}({\vec y},t) $,
which satisfy the standard Poisson brackets
\begin{equation}
\left\{{ A}_{iC}({\vec x},t), \ \Pi^j{}_D({\vec y},t)  \right\}_{PB}= 
\delta^3( {\vec x}-{\vec y}) \delta^j{}_i \delta_{CD}, \label{PB}
\end{equation}
at equal times. Also, we infer from (\ref{HAMAC}) that  $A_{0C}(x)$  
 are 
Lagrange multipliers  leading to the Gauss law constraints
${\cal G}_C=\left( {\cal D}_i E^i \right)_C \approx 0 $, where
${\cal G}_C$
is such that ${\cal G}={\cal D}_i E^i = {\cal G}_C \ T^C $. In particular
\begin{equation}
{\cal G}_C= \partial_i E^i{}_C + f^{BD}{}_C A_{iB} E^i{}_D.
\label{GLC}
\end{equation}

Applying the standard Dirac
method for constrained systems \cite{DIRAC,HT}, we can verify that
there are no secondary constraints.  Also, one obtains  that ${\cal  
G}_C$  are
first-class constraints,  which generate gauge transformations on the
three-dimensional 
hypersurface. A compact way to show these properties is by introducing
the averaged constraints
\begin{equation}
{\cal G}_\Lambda = \int d^3x \ \Lambda^C {\cal G}_C, \label{AVCONST}
\end{equation}
where $\Lambda^C({\vec x}) $ are arbitrary functions of position.  
In this way,
the Poisson brackets of the constraints turn out to be
\begin{equation}
\left\{{\cal G}_\Lambda, \ {\cal G}_\Theta \right\}_{PB}=
{\cal G}_{\Lambda \times \Theta}, \label{CPBA}
\end{equation}
where 
$({\Lambda \times \Theta})_A= 2 f^{BC}{}_{A} \Lambda_B \Theta_C $.  
The group
indices are lowered and raised with the metric $h^{BC}$ together with its
inverse $h_{BC}$. The fact that the constraints ${\cal G}_A$  
generate gauge
transformations can be seen from the following calculation
\begin{equation}
\delta_\Lambda A_{iC}({\vec x},t):=\left\{ A_{iC}({\vec x},t),\  
{\cal G}_{\delta \Lambda } \right\}_{PB} = -({\cal D}_i \delta  
\Lambda )_C,
\label{HGT}
\end{equation}
which reproduces the spatial part of the transformation (\ref{DELCON}). 
The precise  relation between the gauge symmetries obtained  in the 
Hamiltonian formulation and those appearing  in the Lagrangian  
formulation can
be found in 
Ref. \cite{HTZ1}.

The canonical quantization of the Hamiltonian version of a gauge theory
follows the standard steps:

(i)\ The canonical variables are promoted
to the range of hermitian operators: 
$A_{iC} \rightarrow {\hat A}_{iC }$, \ 
$E_{iC} \rightarrow {\hat E}_{iC }$. 

(ii)\ The Pois\-son brack\-ets al\-ge\-bra
is turned into a com\-mu\-ta\-tor (anti\-com\-mutator) algebra,  
according to
the statistic of the involved fields. In our case the resulting  
commutators
are 
\begin{equation}
\left[{\hat A}_{iC}({\vec x},t), \ {\hat E}_{jD}({\vec y},t)  \right]
=i \hbar  
\delta^3( {\vec x}-{\vec y}) \delta_{ij} \delta_{CD}, \label{CR}
\end{equation}
with all others been zero.

(iii)\ A possible way of realizing the above  algebra  is in the  
``coordinate
'' 
representation, where
\begin{equation}
{\hat A}_{iC}:= A_{iC}({\vec x},t), \quad {\hat E}_{iC}:=i{\hbar }
 {\delta \over \delta A_{iC} ({\vec x},t)}.  
\label{REALOP}
\end{equation}
The wave function of the system is then a functional of the
connection: $\Psi(A)$ and we can calculate the action of any  
operator upon it.
The basic functional derivative is defined by
\begin{equation}
{\delta  A_{iC}({\vec x},t) \over \delta A_{jD}({\vec y},t)} =  
\delta^3({\vec
x}-{\vec y}) \delta ^j{}_i
\delta ^D{}_C.
\label{FUNDER}
\end{equation}

(iv)\ The realization of the quantum operators may lead to ordering   
problems. In particular, the first-class
constraints  ${\hat {\cal G}}_C$  must be realized as
operators which correctly close under commutation, in order to have a 
consistent theory. If it not possible to do so, we say that an anomaly
appears. The first-class constraints are subsequently 
imposed as null operator conditions upon the  wave functions  
representing the
physical states 
\begin{equation}
{\hat {\cal G}}_C \ \Psi(A)=0. \label{DIRCOND}
\end{equation}
In our case, the operator expression for the Gauss law constraint
is
\begin{equation}
{\hat {\cal G}}_\Theta =i{\hbar} 
\int d^3x \ ({\cal D}_i \Theta )_C \ {\delta{}  
\over \delta A_{iC}} \label{OPGL}
\end{equation}
and the condition (\ref{DIRCOND}) simply means that the physical  wave
functions 
must be  gauge invariant functionals of the connection.

(v)\ There must be a well defined scalar product which allows for  the
realization of hermitian operators together with  the calculation of
 the
transition probabilities that permits us  to make predictions regarding
the observables of the theory.

(vi)\ Finally, the dynamics  is contained in 
the wave function $\Psi(A,t) $, which satisfies the Schroedinger
equation
\begin{equation}
i \hbar {\partial \Psi(A,t) \over \partial t}= \left( \int d^3x  \  
{1\over2}
\left( {\hat E}^2 + {\hat B}^2 \right) \right) \Psi(A,t). \label{SCHEQ} 
\end{equation}

A corresponding  Hamiltonian formulation  can be also developed for
Chern-Simons theories,
which  are  connection theories defined in a space-time having an   
odd number
of  dimensions.
These topological field theories are  defined by a Lagrangian  
density  given by
the
$2n-	1$ form
$\Omega_{CS}$ defined  by  $d \Omega_{CS}= Tr ( F^n)$. Here,  $F$  
is the 2-form
corresponding to
$F={1\over 2} F_{\mu\nu}  dx^{\mu} dx^{\nu} $, where $F_{\mu\nu}$  
is the field
strength given in Eq. (\ref{FMUNU}). In the case of $2+1$ dimensions, the
Lagrangian action for the Chern-Simons theory is
\begin{equation}
S=  {1\over 2} Tr  \int_{\cal M} \left( dA - {2\over 3} A \wedge A   
\right )
\wedge A,
\label{CSAC}
\end{equation}
where the integration is made over  a  three-dimensional manifold  
${\cal M}$
and
$A=A_{\mu}dx^{\mu}$ is the connection 1-form.

\section*{3.  Yang-Mills Theories with a Connection defined  over a  
Super Lie
Algebra}

Super Lie algebras are a special case of graded Lie algebras, which  
appear in
the description of supersymmetry.   This new type of  symmetry   
arises when
attempting to
provide a unified description of bosons and fermions. In this case,
the  basic object  will be a  multiplet
incorporating  both kind of fields.  The allowed ``rotations''  of  
the theory
will mix  the  components of such multiplet: i.e. bosons and  
fermions.  In this

section we will present only a brief 
review of  some underlying
ideas in supersymmetry, which are relevant to the construction of
supersymmetric
connection theories \cite{SUSY}.

\subsection*{3.1. Grassmann Numbers}

Let us  start from   the ``classical'' version of supersymmetric  
theories,
which requires the introduction of Grassmann numbers to represent
the algebraic properties of fermionic fields. The simplest way to 
motivate these numbers is by  starting from the quantum description
of two independent fermionic operators $f$ and $g$, defined by
the following anticommutators
\begin{equation}
\{f, \ f^{\dagger} \}=1,\ \ \  \{f, \ f \}=2f^2=0, \ \ \ \
 \{f^{\dagger}, \ g^{\dagger} \}=0,
 \ \ \  \{ f, \ g^{\dagger} \}=0,\label{FCR}
\end{equation}
with analogous  expressions  obtained  interchanging $f$ and $g$.
Consider  a realization of the above anticommutation relations
in terms of coordinates and derivatives in a manner analogous to the
holomorphic
representation for the standard harmonic oscillator.To this end, let us
introduce two
independent  coordinates
$\theta$ and $\eta$  and define the following  realization of the above
operators  
\begin{eqnarray}
<\theta \ \eta|f^{\dagger}&=&
\theta<\theta \ \eta| \quad
<\theta \ \eta|f={d \over d\theta}<\theta \ \eta|, \nonumber\\
<\theta \ \eta|g^{\dagger} &=& \eta<\eta \  \theta|,  \quad 
<\theta \ \eta|g={d \over d\eta}<\theta \ \eta|. 
\label{ROFCR}
\end{eqnarray}
The third relation in Eq.(\ref{FCR}) requires 
$\theta \eta+\eta \theta=0$, while the second relation demands
that each of the above numbers must have zero square, i.e. $\theta^2=0
=\eta^2$. 
The last relation in Eq.(\ref{FCR}) says that ${d\over d\theta}\eta
+\eta {d\over d\theta}=0$, which require that the derivative operator
anticommutes with the coordinates. 
The above realization of the fermionic operators  must be supplemented
with a scalar product which guarantees that in fact
$\theta^\dagger={d\over d\theta}$. For the  independent   
coordinates $\theta,
\theta^*$,  for example, such scalar product is given by
\begin{equation}
\left( \Psi , \Delta \right)= \int  d\theta d\theta^* \   \Psi^* \
e^{\theta^* \theta} \  \Delta,
\label{PEGV}
\end{equation}
where the integration over the Grassmann  variables  is defined  by
\begin{equation}
\int  d \theta  =0, \quad  \int d\theta \ \theta = 1,
\label{INTGN}
\end{equation}
and analogously for the  complex conjugated variable $\theta ^*$.

The ``numbers'' $\theta, \theta^* $ and $\eta, \eta^*$
that satisfy the above properties are called odd Grassmann numbers and 
provide a ``classical'' description of fermionic degrees of  
freedom, in the
same
way as complex numbers allow for the realization of bosonic 
degrees of freedom. The product $\theta \eta= - \eta \theta$ is  
called an even
Grassmann
number because it commutes either with $\theta$ or $\eta$ and also  
has zero 
square. Thus, the set of all commuting numbers is augmented from the
complex numbers to the set of even Grassmann numbers. The reader is
encouraged to look into Ref. \cite{DEWIT} for a detailed
description of the properties of Grassmann numbers.

In this way, a unified description of bosonic and fermionic degrees
of freedom will start from a multiplet, say, 
\begin{equation}
|u>=[q^1,q^2, \dots,q^m;\theta^1,\theta^2, \dots, \theta^n]^T,  
\label{BFM}
\end{equation}
which contains bosonic degrees of freedom $q^i, i=1,\dots m$, 
described by   
even Grassmann numbers (complex numbers in particular), 
and fermionic degrees of freedom
$\theta^\alpha, \alpha=1,\dots, n$ described by  odd Grassmann 
numbers. The superindex $T$ denotes standard transposition, in such  
a way
that the array (\ref{BFM}) is a column.

\subsection*{3.2. Supermatrices}

The natural operators acting on the state $|u>$ are linear  
transformations $M$
that produce a ``rotated''  state  $|v>=M|u>$,  which  components 
preserve the even/odd character of each entry in the
multiplet. Such object is the  $(m+n)\times (m+n)$ array
\begin{equation}
M=\pmatrix {A & B \cr C & D},\label{SM}
\end{equation}
where the corresponding blocks have the following properties: (i) \ 
$A$ and $D$ are respectively $m\times m$ and  $n\times n$ matrices
with all entries been even Grassmann numbers. (ii) \   
$B$ and $C$ are respectively $m\times n$ and  $n\times m$ matrices
with all entries been odd Grassmann numbers. The array $M$ is called  
a supermatrix and the corresponding action upon the state $|u>$ is  
called a
supersymmetry transformation.

Let us observe that the addition and the
multiplication of two supermatrices ( with both 
operations defined in the standard way)
is again a supermatrix. Since the unit matrix ${\cal I}$ is also a
supermatrix,
the inverse of a supermatrix can also be defined in the usual manner.

An important topic in the study of supersymmetry  is the
construction of the supermatrix  invariants under similarity  
transformations,
which
generalize the basic concepts of the trace and the determinant in 
classical linear algebra. The supertrace, denoted by $Str$, is defined
by
\begin{equation}
StrM:=TrA-TrD,\label{STR}
\end{equation}
where the relative  minus sign is required in order to guarantee the
cyclic property $Str(M_1M_2)=Str(M_2M_1)$, under the presence of
odd Grassmann numbers in each supermatrix. The definition of the
superdeterminant, denoted by Sdet, follows the same  pattern of the  
 classical
case and it is given by
\begin{equation}
SdetM:= exp(Str(lnM)).\label{SDET}
\end{equation}
This expression can be written in infinitesimal form as
\begin{equation}
\delta ln( Sdet M)=Str(M^{-1} \delta M),\label{DSDET} 
\end{equation}
subject to the boundary condition $Sdet {\cal I}=1$. The equation  
(\ref{DSDET})
is a condensed way of writing the partial 
derivatives of the superdeterminant  with respect to the  
supermatrix entries,
in terms  of the inverse supermatrix $M^{-1}$. The   
superdeterminant satisfies

$Sdet(M_1 M_2)= Sdet(M_1) Sdet(M_2)$. Using these properties one  
can find
the general expressions
\begin{equation}
Sdet(M)={det(A-BD^{-1}C) \over detD}={detA \over det(D-CA^{-1}B)},
\label{SDETM}
\end{equation}
where we assume the existence of $A^{-1}$ and  $D^{-1}$. 
The determinants appearing in Eq. (\ref{SDETM}) are defined in the  
usual way
because they involve only even (commuting) Grassmann numbers. The
equivalence among the two forms of writing the superdeterminant
in Eq.(\ref{SDETM}) is proved in Ref. \cite{BHF}.

\subsection*{3.3. The Superconnection}    

Motivated by the above discussion, it is possible to construct
connection theories where  $A_{\mu}$ is extended  to  be
a $(m+n)\times (m+n)$ supermatrix 
\begin{equation}
A_\mu= A_{\mu C} T^C + \psi_{\mu c} S^c, \label{SUPCON} 
\end{equation}
which incorporates bosonic ($A_{\mu C}$)
and fermionic ($\psi_{\mu c}$) gauge fields that are  
even and odd Grassmann functions of position, respectively. The  
generators 
$T^C$ and $S^c$ are purely numerical (complex) matrices 
which have the generic
structure
\begin{equation}
T=\pmatrix {a & 0 \cr 0 & d},\quad S=\pmatrix {0 & b \cr c & 0},
\label{SMG}
\end{equation}
in the same block form of Eq.(\ref{SM}). They are labeled as
even and odd generators, respectively. 
Under an infinitesimal transformation generated by the supermatrix
$\delta \Theta=\delta {\Theta}_C T^C+ \delta {\Theta}_c S^c$,
the connection transforms in the form given in Eq.(\ref{DELCON}). 
In order that this transformation be closed in the algebra generated
by $T^C, S^c$ it is necessary to require the following (anti)commutation
relations among the generators
\begin{equation}
\left[T^A, \ T^B \right]= f^{AB}{}_C \ T^C,\quad
\left[T^A, \ S^b \right]= g^{Ab}{}_c \ S^c,\quad
\left\{S^a, \ S^b \right\}= h^{ab }{}_C \ T^C. \label{SLA}
\end{equation}
The structure (\ref{SLA}) is called a super Lie algebra and  it is 
characterized by the presence of anticommutators among the fermionic
generators besides the standard commutators for the remaining cases.
The anticommutators  are induced from the basic commutator in  Eq.
(\ref{DELCON})
due to the anticommuting properties of the fermionic components of the
multiplet under consideration. We must emphasize that  the  
functions which
multiply an odd generator in
each multiplet , like $\psi_{\mu c}$ or $\delta \Theta_c$, for  
example,  are
fermionic in
character , so that they anticommute among each other, being  
realized  by odd
Grassmann
functions of position.   The quantities $f^{AB}{}_C, g^{Ab}{}_c,  
h^{ab }{}_C $
are the numerical structure constants of the super Lie algebra.

The field strength $F_{\mu\nu}$ is now a supermatrix and  together  
with the
covariant derivative
${\cal D}_\mu$ retain their definitions given in Eqs. 
(\ref{FMUNU}) and (\ref{COVDER}) respectively. Finally, 
the action for a pure Yang Mills  supersymmetric  theory is given by 
an equation analogous to (\ref{COVACT}),
\begin{equation}
S= \int  d^4 x \  {\cal L} = \int d^4 x \ {1\over 4} Str
(F_{\mu\nu}F^{\mu\nu}), 
\label{SCOVACT}
\end{equation}
where the trace has been replaced by the supertrace. The canonical
analysis and quantization of this theory follows analogous steps
to the standard case, except that now fermions have been included
in the connection.

This ideas can be  extended to the construction of supersymmetric  
Chern-Simons
theories.
The  formulation of  gauge theories where the connection is defined  
over a
super Lie algebra  encompasses some interesting cases like
$2+1$ dimensional  supergravity with cosmological constant   
\cite{CVMAN} \ and
also  the standard supergravity in $3+1$ dimensions,  in terms of   
the Ashtekar
variables \cite{UGGAOBPU}.  The case of global supersymmetry is not  
included in
this way.

\section*{4. The Loop approach to gauge theories}

The description of gauge theories only  in terms of gauge invariant  
 objects,
like   the Wilson loops for example,  was initiated by the work of  
Mandelstam
\cite{MAND}. Other earlier references on the subject are
 \cite{MAMI}, \cite{POLYA}, \cite{NAMB}, \cite{TTWU}.
The loop  representation, which is a quantum  Hamiltonian  
representation of
gauge theories in terms of closed curves (loops)  was subsequently  
introduced
in Refs. \cite{GAMB01}, \cite{GAMB02}.
Roughly speaking, the dynamical variables in this method are 
the generalization of the Aharanov-Bohm phase of  electrodynamics to the
general non-abelian case: the Wilson loops.  In the abelian case,  
this phase is
basically the integral 
of the connection around a closed loop and it is a gauge invariant  
object.
In this way, one considers all possible Wilson loops in a three  
dimensional
surface of constant time,  as the degrees of freedom of the system.  
Thus, one
shifts from the space of connections to the space of loops so that   
the wave
function of the system, originally  a functional
of the connection,  turns out to be a functional of loops. Besides  
allowing for
 the construction
of gauge invariant objects, loops provide a natural geometric  
framework to
describe gauge
theories and gravitation. This approach is a non-local description of the
dynamics and it is well
suited to the discussion of non-perturbative effects.
One of the central problems in this method  is to find and  
independent set of
loop-space  degrees of freedom, because as we will see in the sequel, the
loop-space variables 
form an  overcomplete set.

A very nice example of the properties of this formulation in the case
of standard gauge theories is given in a paper by Brugmann \cite{BRUG}
where the lattice gauge theory for the group $SU(2)$ is completely  
solved 
within this scheme, and compared very favorably with alternative  
solutions.
In  the author's words: \break  `` the eigenvalue problem for the  
Hamiltonian
of $SU(2)$ lattice
gauge theory  is formulated in the loop representation, which is  
based on the
fact that the
physical Hilbert space can be spanned by states which are labeled  
by loops.
Since
the inner product between loop states can be calculated analytically, the
eigenvalue
problem is expressed in terms of vector components and matrix  
elements with
respect to the
loop basis. A small-scale numerical computation in $2+1$   
dimensions yields
results which agree with results obtained from other methods. ''

Another success of the loop space approach can be found in its  
application to
Quantum Gravity  \cite {ROVSM}, \cite
{GAMB1}. The use of the Ashtekar variables  \cite{ASH}, allows
to rewrite Einstein gravity  as a connection theory, in terms of a  
selfdual
connection which
lives in the Lie algebra of $SU(2)$,  plus some reality conditions. The
structure of  the constraints in such formulation is that of the  
corresponding

gauge theory plus some extra constraints related  to the
invariance of  gravity under  diffeomorphism in $3+1$ dimensions.   
Using the
loop 
space approach, it has been possible for the first time to find
explicit solutions to the diffeomorphism constraints of the complexified
theory in terms of functionals of knots. The reality conditions,  
necessary to
get back to real Einstein gravity,
subsequently provide a definition of 
the required scalar product  that leads to an interpretation of the  
quantized
theory.  As we will see in the sequel, loop states automatically  
solve the
local $SU(2)$ gauge invariance of the theory. In this way, the three
dimensional diffeomorphism invariance of Einstein gravity can be  
realized by
labeling the states by knot classes, which are just the equivalence  
 classes of
loops under diffeomorphism. Finally, there remains only one constraint
to solve, which is satisfied in terms of  superposition of knot states.

In this  work  we present only a brief
introduction to the loop space approach of gauge theories. Detailed  
versions,
including
applications to quantum gravity, 
can be found in Refs. \cite{GAMB2}, \cite{GAMB3}, \cite{BRUG1},   
\cite{LOLL},
\cite{PULLIN}.

\subsection*{4.1. Open paths, loops and basic dynamical variables}

\subsubsection*{Definitions}

Our discussion will start from  a space-time 
of  the form $\Sigma \times {\cal R}$,  with  $\Sigma$ being  a three
dimensional surface of constant time, where we have defined
the Hamiltonian version of our gauge theory. All curves  to be
handled  in the sequel will lay on this surface.

Let us consider  a  smooth and continuous function $\eta : s \rightarrow
 \Sigma $, where $s$  is a parameter  in the  real interval
$s_1 \leq s \leq s_2$. An open path on the surface,
joining the points $A$ (initial) and $B$ 
(final point) of it, is defined  by 
$\eta_{A}{}^B(s)=\eta_{s_1}{}^{s_2}(s)= \eta$,  
with $ \eta(s_1)=A$ and $ \eta(s_2)=B$. 
We will use greek letters
in the middle of the alphabet to denote open paths. The choice to  
move along 
 the path, either from $A$ to $B$ or viceversa,  defines its 
orientation . Given any oriented path $\eta_{A}{}^{B}(s)$, its
inverse $(\eta^{-1})_{B}{}^A(t)$ is the  original path traversed
in the inverse sense, with $t$ being the corresponding parameter. 

A loop, or a closed path, is an open path where the initial and final 
points coincide, i.e. such that $A=B$. They will be denoted by  
greek letters
from the beginning of the alphabet: $\alpha(s)$, $\beta(s)$, for  
example.
Again, the sense of traversing  the loop defines its orientation.  
In this
way, the inverse of any loop is the same loop run over  in the opposite
sense. 
Finally, it will prove convenient in the sequel to introduce the idea
of a multiloop ${\tilde \gamma}$ as a collection of loops $\alpha_1, 
\alpha_2, \dots, \alpha_p$, which will be denoted by
${\tilde \gamma}=\alpha_1 \cup \alpha_2 \cup  \dots \cup \alpha_p$. 
A multiloop is
a direct product of loops and does not involve any composition property.

\subsubsection*{Composition properties}

It is convenient to define the multiplication  of paths, which we denote
by $ \circ $. Two open paths $\eta_A{}^B$ and $\sigma_C{}^D$
can  be multiplied only  if either (i) $A=D$, or (ii) $C=B$. In the  
first case,

the composed path, called $(\sigma \ \circ \ \eta)_C{}^B$ , is the  
result of
going  first  through the path $ \sigma $ and subsequently through   
the path
$\eta $, in that order. In the second situation we obtain the path
$(\eta \ \circ \ \sigma)_A{}^D $. The multiplication 
$\eta_A{}^B \ \circ \ (\eta^{-1})_B{}^A$ produces what is called a  
tree, 
which amounts to start from  
point $A$ and going back to it, without enclosing any area on  
$\Sigma$. Any two
open paths or loops differing by a collection of trees, are  
considered to be
equivalent.

Two oriented loops, $\alpha$  and $\beta$ can be multiplied only  
when they 
have a point of intersection, say $C$. Then, the loop
$\alpha \ \circ \ \beta$ is obtained starting from the common point
$C$, going   first through the loop $\alpha$ to end up in $C$,  
subsequently
going through the loop $\beta$, 
to finally end up again at the intersection point. In each case, the
sense of travelling is defined by the orientation of the loops to be
multiplied.
{}From now on we will
consider that all loops are based on a
fixed point on $\Sigma$, so that they always have at least this point
in common. This loop space has the structure of a semigroup.

\subsubsection*{Parallel transport matrix}

To any open path $\eta \in \Sigma$,  we associate the group element
\begin{equation}
U( {\eta}_{s_1}{}^{s_2})= P \ exp \ \int_{s_1}^{s_2} ds \ {\dot  
\eta}^i(s)
\ A_i(s), \label{TPM}
\end{equation}
which is the non-abelian generalization of the Aharanov-Bohm phase  
factor.
Here,
${\dot \eta}^i(s)$ denotes the tangent vector of the path $\eta$ and
$P$ stands  for the path-ordering operator defined as
\begin{eqnarray}
&&P \ exp \ \int_{s_1}^{s_2} dt \ M(t) = 1 + \int_{s_1}^{s_2} dt \ M(t) 
+ \int_{s_1}^{s_2} dt \ \int_{t}^{s_2} dt_1 \ M(t) M(t_1) \nonumber \\
&&+ \int_{s_1}^{s_2} dt \ \int_{t}^{s_2} dt_1 \ \int_{t_1}^{s_2} dt_2 \
\dots \quad  M(t) M(t_1) M(t_2) \dots  := V(s_1, s_2),
\label{POO}
\end{eqnarray}
where $M(t)$ denotes any element of the  corresponding Lie algebra,  
valued
along the
path $\eta$. Let us 
emphasize that, in general, $\left[ M(t_1), \ M(t_2) \right] \neq 0$
for arbitrary points in the path. The ordering in
(\ref{POO}) is defined  by $s_1  < t < t_1 < t_2 < \dots  < s_2$. An
alternative  way of
writing  the generic  ordered integral is in the form
\begin{equation}
 \int_{s_1}^{s_2} dt_2 \ \int_{s_1}^{t_2} dt_1 \ \int_{s_1}^{t_1} dt \
\dots \quad  M(t) M(t_1) M(t_2) \dots \ .
\label{POO1}
\end{equation}
The parallel transport matrix  (\ref{TPM}) is also known as the  
integrated
connection along the  corresponding path. As a function of the end  
points, the
group element $V(s_1, s_2)$ satisfies
the properties
\begin{equation}
{ \partial V(s_1, s_2) \over \partial s_2}= V(s_1, s_2) M(s_2),\quad
{ \partial V(s_1, s_2) \over \partial s_1}= - M(s_1) V(s_1, s_2), 
\label{DTPM}
\end{equation}
which can be directly seen from Eq. (\ref{POO1}) and  Eq. (\ref{POO})
respectively.
Also we have the composition property
\begin{equation}
U( {\eta}_{s_1}{}^{s_2}) \  U( {\sigma}_{s_2}{}^{s_3})=
U( ({\eta \ \circ  \ \sigma})_{s_1}{}^{s_3}),
\label{COIC}
\end{equation}
provided the final point of the path $\eta$ coincides with the 
initial point of the path $\sigma$. In this way, both paths
can be composed as indicated in the RHS of Eq. (\ref{COIC}).
The composition appearing in the LHS of  Eq.(\ref{COIC}) corresponds
to the group multiplication.
Finally, under  non-abelian finite gauge transformations, 
(generated by the group element $g(x)$),
$A_i \rightarrow {\tilde A}_i= g A_i g^{-1}- g \partial_i g^{-1}$,
the integrated connection transforms as
\begin{equation}
U( {\eta}_{s_1}{}^{s_2})\rightarrow {\tilde U}( {\eta}_{s_1}{}^{s_2})=
g(s_1) U( {\eta}_{s_1}{}^{s_2}) g^{-1}(s_2).
\label{GTIC}
\end{equation}

\subsubsection*{Holonomies and Wilson loops}   

The final goal of this formulation is to define strictly gauge
invariant objects, which will constitute the appropriate variables
to formulate the theory.  To this end, let us consider
 the holonomy $U(\alpha)$ associated to the loop $\alpha$, which is  
just the integrated connection (\ref{TPM}) around the loop,
\begin{equation}
U(\alpha)= P \ exp \oint_\alpha ds {\dot \alpha}^i(s) A_i(s). 
\label{HOLO}
\end{equation}
This object is still not gauge invariant, but transforms according to
(\ref{GTIC}) with the corresponding group elements evaluated at   
different
values of the parameter, which nevertheless describe the same point on
$\Sigma$: the chosen initial $(s=s_1)$  and final  point  $(s=s_2)  
$ of the
loop. Group elements must
be single valued on $\Sigma$, so that $ g^{-1}(s_2)=  
(g(s_1))^{-1}$. In this 
way,   $Tr U(\alpha)$ is indeed a gauge invariant complex number.  
This is the
original Wilson loop, whose definition has been generalized to include
insertions of the gauge-covariant chromoelectric field, as a way to 
incorporate
the variable conjugated to the connection in a gauge invariant  
manner. Then, a
tower
of Wilson loops can  be constructed as follows
\begin{eqnarray}
T^0(\alpha):&=& Tr \ U(\alpha) = h(\alpha, A), \nonumber \\
T^{i}(\alpha):&=&Tr \left[U({\alpha}_{s_0}{}^{s_1}) E^i(s_1) 
U({\alpha}_{s_1}{}^{s_0}) \right], \nonumber \\
T^{ij}(\alpha):&=& Tr \left[U({\alpha}_{s_0}{}^{s_1}) E^i(s_1) 
U({\alpha}_{s_1}{}^{s_2}) E^j(s_2) 
U({\alpha}_{s_2}{}^{s_0}) \right], \dots ,
\label{TWL}
\end{eqnarray}
where $s_0$ parametrizes the origin of the loop $\alpha$.
Each of the numbers in (\ref{TWL}) is a gauge invariant quantity  
and we will
denote
them generically by $ T^0, T^1, T^2, \dots$  according to the number of
insertions introduced. These objects constitute the fundamental  
dynamical variables used  to formulate the gauge theory in this  
approach.

The original symplectic structure (\ref{CR}) of the gauge theory   will 
induce a Poisson brackets structure  among the Wilson loops of 
Eq.(\ref{TWL}). In a very sketchy form, these Poisson brackets will
be of the form 
\begin{equation}
\left \{ T^m, \ T^n \right\}_{PB} = \sum_{k=1}^{m+n-1} 
C_{k}T^{k}.
\label{PBWL}
\end{equation}
The result in the RHS  will involve some recomposition and rerouting
of the initial loops appearing in the LHS. The details are given in Refs.
\cite{GAMB02}, \cite{ROVSM}.

\subsection*{4.2. The loop space}

The dynamical  variables defined above , i.e. the Wilson loops,    
have support
on the space of
closed loops on $\Sigma$, which now replaces the configuration  
space associated
to  the
connection representation. To complete the formulation of the gauge  
theory in
the loop
space is still necessary to rewrite the Hamiltonian and the  
constraints in the
new representation.
This can be accomplished through the introduction  of differential  
operators
which act on loop dependent functions: the loop derivative and the  
connection
derivative, for example. The reader
is referred to Ref. \cite{GAMB3} for the details.

Accordingly, when quantizing the theory,
the corresponding operators $T^n \rightarrow {\hat T}^n$
will have also  support on the space of  loops and they will
be constructed as operators acting  on a Hilbert space  which now will be
spanned by
functionals of loops. 
An heuristical, but very illuminating way  of looking at some properties
of  the loop space is via the loop-transform introduced in Refs.  
\cite{ROVSM},

\cite{GAMB1}.This transforms provides the change of basis
from the connection representation $\{|A>\}$ to the loop representation
$\{|{\tilde \alpha} > \}$ in the form
\begin{eqnarray}
|{\tilde  \alpha} >&=& \int [dA] \ |A><A|{\tilde \alpha} >, \nonumber \\
<A|{\tilde \alpha} >&=& {\cal K}({\tilde \alpha}, A) := TrU(\alpha_1) 
 TrU(\alpha_2) \dots TrU(\alpha_p),
\label{LT}
\end{eqnarray}
where ${\tilde \alpha}$ is a multiloop with components 
$\alpha_1, \alpha_2, \dots, \alpha_p$. The measure $[dA]$ 
in (\ref{LT}) is  not  presently known and this fact  is what makes this
transform
only an heuristical tool. The wave function of the system $| \Psi >$
can then be projected either  into the connection representation  
$<A| \Psi>$
or  into  the loop representation $<{\tilde \alpha} | \Psi>$, with the
two projections being  related by the loop transform (\ref{LT}).
The change of basis (\ref{LT}) will provide also the representation  
of the
operators acting
in the loop space, starting  from their counterparts in the connection
representation.

Nevertheless, there is an alternative way  to this construction,  
based upon the
formulation
 of the quantum theory directly  in the  loop space. To this end, one
starts from the Poisson brackets relations (\ref{PBWL})  among  the  
Wilson
loops. 
Following the standard procedure, the classical variables are  
promoted into
operators ${\hat T}^n$ and their Poisson brackets are translated into
commutators. A basis in the loop space is  constructed by  first  
defining the
zero-loop
state $|0>$ such that ${\hat T}^n |0> = 0, \  n\geq 1$. An arbitrary
multiloop state $|{\tilde \alpha}>$ is subsequently constructed 
by the action ${\hat  T}^0(\alpha_1) {\hat  T}^0(\alpha_2)\dots {\hat
T}^0(\alpha_p) |0>:=|{\tilde \alpha}>$.
Then, using the commutation relations it is possible to figure out  
the action 
of any operator upon the multiloop states. Ref. \cite{BRUG} 
contains a very nice and explicit application of these ideas.

Being gauge invariant, the Wilson loop variables automatically  
satisfy the
Gauss law constraints
of the theory. In this way, the arbitrary multiloop  quantum states  
constructed
by the action
of the operators ${\hat  T}^0(\alpha_1) {\hat  T}^0(\alpha_2)\dots {\hat
T}^0(\alpha_p)$ on the zero loop state, will be automatically  
annihilated by
the quantum Gauss law constraints.

The equivalence between the connection representation and the loop
representation
of  gauge theories defined over a Lie algebra  was proved in  
Ref.\cite{GILES}
for the group
$SU(N)$. There,  it is given
a  procedure to reconstruct, up to a gauge transformation,  the  
gauge field at
each point  ( written as a complex matrix ) starting from the  
knowledge of all
the Wilson loops
$T^0(\alpha)$ going through that point, together with the  
imposition of the so
called Mandelstam identities, which set  constraints among the
Wilson loops. This construction demonstrates that the Mandelstam  
identities are
a sufficient set
of conditions on the  Wilson loop variables which  guarantee that such
variables represent a local
gauge theory over a Lie algebra.

\section*{5. Overcompleteness of the loop space variables}

Let us summarize what has been done by saying that the loop space 
formulation of classical  gauges theories starts from an infinite  
set $ \{T^n
\}$
of numerical, non-local,
gauge invariant degrees of freedom, defined over  loops in the
three-dimensional surface $\Sigma$. These variables replace those 
of the standard local connection representation,  $E({\vec x},t), A({\vec
x},t)$, which  are
not gauge invariant.

The loop space  degrees of freedom  are basically  traces of group
elements,
which in the sequel  will be thought of as traces of  matrices of a given
dimension.
Now, it is well known that the
Cayley-Hamilton theorem of linear algebra, for example, sets  
relations among
the
traces of powers of a given matrix, which depend on the dimension
of the representation. As we will show in the sequel, a similar mechanism
induces relations among the different  Wilson loops  $T^0 (\alpha)$, 
with
the consequence that   $ \{T^n \}$  is  an overcomplete set of dynamical
variables.
These relations among powers of traces of products of matrices are  
generically
known as Mandelstam identities (MI). Even in the
standard case (i.e.  gauge theories over a Lie algebra) there is  
lacking a
general
procedure to obtain the full set of  MI for arbitrary restricted  
groups. The
situation
in the case  where the connection is valued on  a super Lie algebra  
is still
more open and we  hope that some of the  ideas presented here could  
provide a
starting point  to deal with these problems.

At the quantum level, these MI will generate, via  the loop transform
(\ref{LT})
for example, relations among the  multiloop
states, which  will make this set also overcomplete.

In this way, any loop space formulation of a gauge theory
will require the knowledge of the corresponding complete set of MI, in
order that the true degrees of freedom of the system can  be identified.
The subsequent problem
of  imposing and/or  solving   the  MI is still  highly involved  
and has been
carried over,
for example, in the following cases:  (i)  $SU(2)$  pure lattice gauge
\cite{LOLL1}, \cite{WATSON}. (ii) $2+1$ dimensional Einstein gravity
\cite{RENEL}  and (iii) a specific sector of $2+1$ supergravity  
\cite{URRU1}.

 The observation made above concerning the relation 
between  the MI and the Cayley-Hamilton identities  will be made more
explicit in what follows,  by   formulating an algorithm  to go  
from one type
of
identity  to the other, in  the
generic case of matrices  over the complex numbers.  This procedure  
will be
next 
extended to generate the generic  MI for supermatrices, after having also
extended
the Cayley-Hamilton theorem to this  case.

\subsection*{ 5.1. The Mandelstam identities}

The simplest MI (known as MI of the first class) is just the  
statement of the
cyclic property
of the trace. For any two  complex square matrices $M$ and $N$, we have
\begin{equation}
Tr \left(  M N  \right)=Tr \left(  N M   \right).
\label{CPOT}
\end{equation}
Let us consider the effect of this identity upon the loop space variable
$T^0(\alpha \circ \beta)$
for example. We have  $T^0(\alpha \circ \beta)= Tr  \  U (\alpha  
\circ \beta)
=
Tr \ ( U(\alpha) U(\beta)) =  Tr \ ( U(\beta) U(\alpha) )$, where  
we have used
the composition
property (\ref{COIC}) for the holonomies.  The above chain of  
equalities leads
to the
conclusion
\begin{equation}
T^0(\alpha \circ \beta)= T^0(\beta \circ \alpha).
\label{CMR1}
\end{equation}
When translated into an statement over the quantum loop states,  Eq.
(\ref{CPOT}) implies
\begin{equation}
| \alpha \circ \beta >= | \beta \circ \alpha >,
\label{QMR1}
\end{equation}
which  arises either from  the loop transform (\ref{LT}), or  from the
method of creating
the quantum  loop space  vectors starting from the zero loop state  
$|0>$. The
above constitutes  a very
simple example
of the overcompleteness of  both  the classical loop space  
variables together
with  the
corresponding quantum states.

 All the remaining MI are known as Mandelstam identities of the  
second class
and there
is a hierarchy of them, according to the restrictions satisfied by  
the group
under consideration.
Suppose we are dealing with a group representation in terms  of  
$n\times n$
matrices $M$. There will be what we call a generic MI, related to  
the specific
dimension $n$. If we impose further restrictions upon the representation,
like unit determinant or unitarity, for example,  new MI reflecting  
them have
to be generated.  Most of what we will discuss in these notes has  
to do with
the generic MI and their extension to the supersymmetric case.

\subsubsection*{ The generic Mandelstam identities}

Let us consider the group elements $U(\alpha)$  realized as $n\times n$
matrices
with no further restrictions. The corresponding
generic MI are obtained starting from the identity\cite{GILES}
\begin{equation}
0= \epsilon_{i_1 i_2 \dots  i_n i_{n+1}} \epsilon_{j_1 j_2 \dots   
j_n j_{n+1}}
 = \sum_{P \in S_{n+1}} \ (-1)^{\pi(P)}  \ \delta_{i_1 P(j_1)} \dots
\delta_{i_{n+1} P(j_{n+1})},
\label{EPID}
\end{equation}
where $  i_k, j_k= 1, \dots n, \quad k=1, \dots, n, n+1.$ Here, the  
sum is
over
all permutations
$P$ of the symmetric group $S_{n+1}$ and $\pi(P)$ denotes the  
parity of the
permutation.  The
zero in the LHS of Eq.(\ref{EPID}) comes about  because there must be a
repetition when
distributing $n$ objects among $(n+1)$ places. The second equality  
is just the
determinantal expansion of the
product of two $\epsilon$-symbols.

By saturating the relation (\ref{EPID}) with $n+1$ different  
matrices one
obtains an identity
among products of traces of products of matrices. The resulting  
Mandelstam
identity
can be written as  \cite{GLIOZZI}, \cite{GAMB5}
\begin{equation}
\sum_{Perm(1,2,\dots, n,n+1)} (-1)^{\pi(P)} W(M_1, M_2, \dots, M_n,
M_{n+1})=0,
\label{MI1}
\end{equation}
where
\begin{equation}
W(M_1, M_2, \dots, M_n, M_{n+1})= Tr (M_{a(1)} \dots M_{a(i)}) Tr(  
M_{a(i+1)}
\dots ) \dots,
\label{MI2}
\end{equation}
corresponds to the cycle decomposition  $(a(1) \dots a(i)) ( a(i+1)  
\dots
)\dots $ of the permutation $P$. The generic MI is a function of  
degree one in
each of the $n+1$ different matrices
involved.  As an example, let us consider the $2\times2$ case,  
which leads to
the following
MI among three matrices
\begin{eqnarray}
0&=& Tr (M_1) Tr (M_2)Tr (M_3) + Tr (M_1 M_3 M_2 ) + Tr (M_1 M_2 M_3 )
\nonumber \\
&&- Tr (M_1) Tr (M_2 M_3) - Tr (M_2) Tr (M_1 M_3) - Tr (M_3) Tr  
(M_1 M_2).
\label{MI22}
\end{eqnarray}
The  restriction implied among  the  classical loop variables  is
\begin{eqnarray}
0&=& T^0(\alpha_1) T^0(\alpha_2)T^0(\alpha_3) + T^0(\alpha_1 \circ  
\alpha_3
\circ \alpha_2) +
 T^0(\alpha_1 \circ \alpha_2 \circ \alpha_3) \nonumber \\
&&-T^0 (\alpha_1) T^0 (\alpha_2 \circ  \alpha_3) - T^0 (\alpha_2)  
T^0(\alpha_1
\circ  \alpha_3) - T^0 (\alpha_3) T^0 (\alpha_1 \circ \alpha_2),
\label{CC22}
\end{eqnarray}
which has the following expression in terms of  the quantum loop states
\begin{eqnarray}
 0&=& | \alpha_1 \cup \alpha_2 \cup \alpha_3> + | \alpha_1 \circ  
\alpha_3
\circ
\alpha_2>
+ | \alpha_1 \circ \alpha_2 \circ \alpha_3> \nonumber \\
&&| \alpha_1 \cup \alpha_2 \circ \alpha_3> + | \alpha_2 \cup  
\alpha_1 \circ
\alpha_3>
+ | \alpha_3 \cup \alpha_1 \circ \alpha_2>.
\label{QC22}
\end{eqnarray}

\subsubsection*{  The restricted Mandelstam identities}

To the author's  knowledge, there is no systematic procedure to  
construct the
MI that reflect further restrictions on the group elements. In this  
section
we give a brief description of the method used in Ref. \cite{GAMB5} to
implement the MI for the group  $SU(N)$
realized in terms of  $N\times N$ matrices $M_i$ in the adjoint  
representation.
In this
reference, the basic objects to be considered are 
the numbers ${\cal W}_k(M_1, M_2, \dots M_k)$ which are defined through
the following recurrence relation 
\begin{eqnarray}
{\cal W}_1(M_1)&=& Tr (M_1),\nonumber \\
(k+1) {\cal W}_{k+1}(M_1, M_2, \dots M_{k+1})&=&
{\cal W}_1(M_{k+1}){\cal W}_k(M_1, M_2, \dots M_k)\nonumber \\
&-& {\cal W}_k(M_1M_{k+1}, M_2, \dots M_k) \nonumber \\
&-& \dots - {\cal W}_k(M_1, M_2, \dots M_k M_{k+1}).
\label{GAMRR}
\end{eqnarray}
We have changed the notation with respect to Ref.\cite{GAMB5}, in order
to be consistent with our own conventions.
In this construction, the generic MI is written as 
\begin{equation}
{\cal W}_{N+1}(M_1, M_2, \dots M_{N+1})=0.
\label{GAMMI}
\end{equation}
Unitarity is imposed by demanding
\begin{equation}
{\cal W}_{1}(M^{-1})={\cal W}_{1}(M)^*,  \quad |{\cal W}_{1}(M)|\leq N. 
\label{GAMBU}
\end{equation}
Finally, the unit determinant condition can be written as
${\cal W}_{N}(M, M, \dots M)=1 $. This, in turn, can be reexpressed
in terms of $N+1$ matrices in the form
\begin{equation}
{\cal W}_{N}(M_1 M_{N+1}, M_2 M_{N+1}, \dots M_{N} M_{N+1})=
{\cal W}_{N}(M_1, M_2, \dots M_N).
\label{DET1}
\end{equation}
This last form is useful to produce a geometric
interpretation of the condition of unit determinant in the language
associated with the loop description of this gauge theory.

\subsection*{5.2. The  relation between the Mandelstam and the 
Cayley-Hamilton identities}

In order to motivate the results of this section,  
let us consider an interesting
particular case of Eq.(\ref{MI22}), by setting $M_1=M_2=M, \  
M_3=X$, where 
$X$ is an arbitrary $2\times2$ matrix. Substituting  in  Eq.  
$(\ref{MI22})$
and
using the  property  that $Tr (CX)=0, \  \forall X$,  implies  
$C=0$, we obtain
the matrix relation
\begin{equation}
M^2 - Tr(M) M + det(M){\cal I}_2=0,
\label{CH22}
\end{equation}
where ${\cal I}_2$ is the $2\times 2$ identity matrix. In the above  
we have
used the fact that $  det (M)={1\over 2}\left( {Tr(M)}^2 - Tr(M^2)  
\right)$
for  $2\times 2$
matrices. Next,  we recognize that Eq.(\ref{CH22}) is nothing but the
statement
of the Cayley-Hamilton theorem for the $2\times 2$ matrix $M$. As  
we will show
in 
the sequel, this construction can be generalized to $n\times n$ matrices
and can  also be used in the reversed sense, that is to say,   
starting from
 the identities
arising from the Cayley-Hamilton theorem, an algorithm to obtain   
the generic
MI in the general case can be constructed.
\paragraph*{The Cayley-Hamilton theorem in classical linear  
algebra.} Let us
provide a brief review of this important theorem.  For an arbitrary  
$n\times
n$
complex matrix $M$, the corresponding characteristic polynomial  
$P(x)$ is
defined as
\begin{equation}
P(x)= det \left( x{\cal I}_n - M \right)= x^n + a_1 x^{n-1} + \dots  
 a_n.
\label{POLCAR}
\end{equation}
The Newton equations provide a recursive method to calculate the  
coefficients
$a_i$,
in terms of  traces of   powers of $M$  \cite{USPEN}
\begin{equation}
a_{i+1}= - {1  \over i+1 } \sum_{k=0}^i   t_{i+1-k}  \  a_k,
\label{RRCP}
\end{equation}
where  $t_k = Tr (M^k)$.
An explicit solution of the above  recursion 
relations is
\begin{equation}
a_{i+1}= \sum_{\alpha_0 + \alpha_1 + \dots + \alpha_s=i+1} {  (-1)^{s+1}
t_{\alpha_0}  t_{\alpha_1}\dots
t_{\alpha_s}  \over
(\alpha_0 + \alpha_1 + \dots  + \alpha_s) ( \alpha_1 + \dots +  
\alpha_s) \dots
(\alpha_s)  },
\label{SRRCP}
\end{equation}
where the sum\-ma\-tion is over all the un\-ordered dis\-tinct  
par\-ti\-tions
$(\alpha_0,  \dots,  \alpha_s) $  of $i+1$, with $s+1 $ being  the
total number of terms in each partition.

Extending the complex variable $x$ to a matrix-valued variable, the
Cayley-Hamilton theorem
states that each matrix  annihilates its characteristic polynomial  
\cite{DCHT},
i.e.
\begin{equation}
P(M)= M^n + a_1 M^{n-1} + \dots  a_n {\cal I}_n=0.
\label{CHT}
\end{equation}
Let us emphasize that (\ref{CHT}) is a matrix identity involving $n^2$
numerical relations.
We will refer to the  relations of the type   (\ref{CHT}) as the
Cayley-Hamilton
identities (CHI).

Now we describe the proposed algorithm to obtain the MI starting  
from the CHI
\cite{URRU2}. The basic idea is to produce a sequence of CHI that  
leads to the
MI, incorporating $n+1$ different
matrices and recalling  that the final MI must be an homogeneous  
function of
degree one in
each matrix. The procedure is as follows:  starting from  the   
identities
$P_{M_1}(M_1)=0, \  P_{M_2}(M_2)=0,  \ P_{M_1 + M_2}(M_1 +M_2)=0$  
we construct
\begin{equation}
T_2(M_1, M_2):= P_{M_1 + M_2}(M_1 +M_2)- P_{M_1}(M_1)-P_{M_2}(M_2) =0.
\label{T2}
\end{equation}
We are denoting by $P_M(x)$ the characteristic polynomial  
corresponding to the
matrix $M$
 and also we have  set  $P_M(M):=T_1(M)$.
Every term in (\ref{T2}) is an  homogeneous function of $M_1$ and  
$M_2$ and
the
subtractions
are designed in  such a way  that  both  $M_1$ and $M_2$ appear at  
least once
in
every  term of $ T_2(M_1,
M_2)$. In this way $T_2(M_1=0, M_2)$ and $T_2(M_1, M_2=0) $ are  
identically
zero. Moreover, we consider
that $T_2(M_1, M_2)$ can be fully expanded using the distributive  
property of
both the
trace and the
matrix product with respect to matrix addition. The next step is to  
construct
\begin{eqnarray}
T_3(M_1, M_2, M_3)& := &P_{M_1+ M_2+ M_3}(M_1+ M_2+ M_3)|_{red},   
\nonumber \\
&:=& P_{M_1+ M_2+ M_3}(M_1+ M_2+ M_3)- T_2( M_1, M_2) - T_2( M_1, M_3)
\nonumber \\
&&-
T_2( M_2, M_3)  - T_1 (M_1)- T_1 (M_2)- T_1 (M_3) = 0.
\label{T3}
\end{eqnarray}
Again, $T_3$ is identically zero whenever any of the  $M_i$ is set  
equal to
zero. We have introduced the subscript $|_{red}$ to indicate an  
identity which
has been reduced in such a way
that every matrix involved is present at least once in every term,  
after the
identity is fully
expanded. In other words, the expression $P_{M_1+ M_2+ M_3}(M_1+ M_2+
M_3)|_{red} $
in (\ref{T3})  can be directly constructed by expanding the  
corresponding
characteristic
polynomial and discarding all terms in which any one of the three  
matrices is
missing.

Extending this idea, we  construct reduced
identities of always
increasing order, where we subtract all the lower order identities  
at our
disposal. This procedure leads to 
\begin{eqnarray}
T_k (M_1,   \dots,  M_k)&:=&P_{M_1+ \dots + M_k}(M_1+ \dots +  
M_k)\nonumber \\
&& - \sum_{i < k}
T_i (M_{s_1}, \dots, M_{s_i}),
\label{TK}
\end{eqnarray}
where the sum is carried over all subsets $\{ s_1, \dots, s_i\} $  
of $\{ 1,
\dots, k \}$. The fact that
the char\-ac\-ter\-is\-tic poly\-no\-mial is of or\-der    $n$  \
guar\-an\-tees that  
$T_k (M_1,   \dots,  M_k)$ are identically zero for  $  k \geq  n+1  
$. The
expression
$T_n (M_1,   \dots,  M_n)$ is
homogeneous of degree one in each of the $n$ matrices involved.

The generic MI  is  obtained from $Tr \left( T_n (M_1,   \dots,   
M_n) M_{n+1}
\right)$ as shown with all detail 
in Ref. \cite{URRU2}. The converse statement, which amounts to
recovering  the CHI (\ref{CHT}) starting from the MI (\ref{MI1}),    
is also
proved in general in this work.

Before closing this section let us observe that the immediate   
application of
the above
method to the construction of 
the MI  in  the case where  further restrictions upon the group  
elements are
required,  will work
only if  such restrictions are preserved by  matrix addition  
\cite{GAMB6}. In
particular, the restriction to matrices with
unit determinant cannot be directly  implemented in this way. It  
remains still
an open
problem to work out
a general procedure to construct the MI corresponding to  arbitrarily
restricted
groups.

Nevertheless,  there are two   interesting examples where  a  
restricted MI is
directly 
obtained  from the characteristic polynomial.  These correspond to   
the cases
of the
groups $SL(2, \Re)$ and $SU(2)$,  which are relevant in the study   
of $2+1$ de
Sitter gravity, and $3+1$ Einstein gravity in the Ashtekar
formulation, respectively. Let us consider two of  such  $2\times 2$
matrices,
$M_1$ and $M_2$, with unit determinant. The CHI for the first  
matrix can be
rewritten as
\begin{equation}
M_1 -Tr(M_1) {\cal I}_2 + M_1^{-1}=0,
\label{CHISU}
\end{equation}
after the Eq.  (\ref{CH22}), with $det(M_1)=1$, has been multiplied by $
M_1^{-1}$.
Multiplying further Eq. 
(\ref{CHISU}) by $M_2$ and taking the trace, we end up with the  
restricted MI
\begin{equation}
Tr( M_1 M_2^{-1}) + Tr (M_1 M_2)= Tr(M_1) Tr(M_2),
\label{MISU}
\end{equation}
which involves only   two matrices, instead of three as in the  
generic case
presented
in Eq. (\ref{MI22}).

In the case of  the group   $SL(2, \Re)$ it is also possible to  
exhibit an
example of the drastic reduction of the
independent
loop space variables produced by the MI. Let us consider the  
infinite set of
Wilson loops
of the form $ Tr ( M_1^{p_1} M_2^{q_1} M_1^{p_2} M_2^{q_2} \dots  
M_1^{p_n}
M_2^{q_n}
\dots ) $, for any  integer $p_i, \ q_i$. Using the MI (\ref{MISU})
one can show  that any of such Wilson loops
can be expressed as a function of three traces only: $Tr(M_1), \  
Tr(M_2)$ and
$
Tr(M_1 M_2)$\cite{NEREZE}.
Thus, in the corresponding sector of the theory we will have only three
independent degrees
of freedom. A simple example of such reduction is to consider  
$Tr(M_1^2 M_2)$
for
example. Here
we apply the relation (\ref{MISU}) with $M_1 \rightarrow M_1, \ M_2
\rightarrow
M_1 M_2$
obtaining
\begin{equation}
Tr(M_1^2 M_2) = Tr(M_1) Tr(M_1 M_2) - Tr(M_2).
\label{EXTRR}
\end{equation}

\section*{6. The Cayley-Hamilton theorem for supermatrices}

Our main motivation for the work described in the previous section  
has been
the
possibility
of extending these ideas to the loop space  formulation  of pure  
gauge theories
having  their  connection valued on a super Lie algebra, as it is  
the case
of  supergravity in the
Ashtekar variables formalism \cite{UGGAOBPU}, for example.
In this situation, the
group elements are described by  generic supermatrices as defined in
 Eq. (\ref{SM}). The
knowledge of the  MI for supermatrices  will be  also  relevant to  
the loop
space
formulation of  fully supersymmetric gauge theories.

It is well known that  Grassmann numbers  can be realized
in terms of complex  numerical matrices.
{}From this
point of view one could think  that  the Cayley-Hamilton theorem for
supermatrices
would be  just
a trivial extension of the standard case. Nevertheless, this is not  
the case
for
at least  two reasons: (i) after a  realization of the Grassmann  
numbers in
terms of
numerical matrices  it would not  be possible
to recover a result in terms of the original Grassmann numbers .  
(ii) the
matrix realization  will effectively  augment the size of
the resulting supermatrix , which now would  be completely  
numerical, in such
a
way that the resulting standard characteristic polynomial would be  
also of
higher
degree.

Let us give a precise meaning to the above observations  using the
simplest case
of a  $(1+1)\times (1+1)$ supermatrix
\begin{equation}
M_2 =\pmatrix {p & \eta \cr \theta & q},\label{SM22}
\end{equation}
where $p, \ q$ are even Grassmann numbers while $\eta, \ \theta$ are odd
Grassmann numbers  such that
$\eta^2=0=\theta^2, \ \eta \theta= - \theta \eta $.  The minimum  
size for the

realization of  these Grassmann numbers  corresponds  to $4\times 4$
gamma-matrices, like
$\eta= \gamma_0 + \gamma_1, \  \theta= \gamma_2 + i \gamma_3$ in  
the signature
$ (+, -, -, -)$, for example. In this way, $M_2$ will be realized as an
$8\times 8$ numerical
matrix and, consequently, the resulting characteristic polynomial  
will be  a
numerical
polynomial of
degree $8$. Clearly, it will not be possible to rewrite the numerical
coefficients of the
polynomial in terms of the original Grassmann numbers.  
Nevertheless,  it is
possible
to find a characteristic polynomial  of
degree $2$ for this case,   which  is given by
\begin{equation}
P_2(x)= (q-p) x^2 - (q^2-p^2 - 2 \eta \theta) x + ( (q-p) qp - (q +  
p) \eta
\theta).
\label{PCSM2}
\end{equation}
We can verify,  by direct substitution,  that  the above   
polynomial satisfies
$P_2(M_2)=0$. In this simple case,
the polynomial (\ref{PCSM2})
can be  constructed just by solving this condition. When $(q-p)^2  
\neq 0$, the
polynomial
(\ref{PCSM2}) can be redefined in the monic form of  
Eq.(\ref{POLCAR}) with the
choices $a_1=q+p - { 2  \over {q-p}} \eta \theta $ and $a_2=  qp -{  
q+p \over
{q-p}} \eta \theta $.  Thus, $a_1$ generalizes
$ Tr (M_2) = p+q$,  while $a_2$ generalizes  $det(M_2) = qp- \eta
\theta$  corresponding  to the case where
all entries in $M_2$ would be  complex numbers. Here we also see that the
standard
determinant is not well defined for supermatrices: we  would have  
to make a
choice among all
possibilities $qp - (A\eta \theta + (1-A) \theta \eta)$, for  
arbitrary $A$.  In
this way,  the
heuristical result  (\ref{PCSM2})
certainly motivates the search for a general procedure to construct  
such null
polynomials,
leading to the construction  of  the Cayley-Hamilton theorem for  
supermatrices.
This is an interesting
problem in its own, besides the  possible applications to the loop space
formulation
of  gauge  theories involving supersymmetry.

\subsection* { 6.1. The characteristic and null polynomials for  
supermatrices}

The first step in this direction  is to  provide  a definition of the
characteristic polynomial,
which extends the
standard one given in  Eq.  (\ref{POLCAR}). This problem is  
certainly related
to the
eigenvalue problem of a supermatrix, which is discussed in Ref.  
\cite{JAP1}.
The eigenvalues of an $(m+n) \times (m+n)$ supermatrix $M$  are  
even Grassmann
numbers which can be of  two types: (i) first-class eigenvalues,   
$\lambda_i,
\
i=1, \dots, m$ , whose eigenvectors
are of the form $[ E_1, \dots, E_m, O_1, \dots, O_n]^T$, where  
$E_k$ denotes 
even Grassmann numbers, while $O_l $ labels  odd  Grassmann  
numbers. (ii)
second-class
eigenvalues, ${\bar \lambda}_\alpha, \ \alpha=1, \dots, n$ with  
corresponding
eigenvectors
of the form $[ O_1, \dots, O_m, E_1, \dots, E_n]^T$.
The  characteristic polynomial
will  certainly read
\begin{equation}
P(x) = (x-\lambda_1) \dots (x-\lambda_m) (x-{\bar \lambda}_1) \dots  
 (x-{\bar
\lambda}_n),
\label{SPOLCARE}
\end{equation}
in terms of the eigenvalues of the supermatrix .
When written in the form of Eq. (\ref{POLCAR}), the explicit  
expressions for
the
coefficients $a_i$
will  be   given again  by Eq. (\ref{SRRCP}) with $t_k= \lambda_1^k  
+ \dots +
\lambda_m^k
+ {\bar \lambda}_1^k + \dots +  {\bar \lambda}_n^k $ .  A  problem
now arises
if  we want to rewrite such coefficients in terms of the  
supermatrix itself :
the appropriate
invariants for this case  are not  traces of powers of the  
supermatrix, but  
supertraces of
powers of the supermatrix instead. The latter are given by
$Str (M^k)= \lambda_1^k + \dots +  \lambda_m^k  - {\bar  
\lambda}_1^k - \dots -
{\bar \lambda}_n^k $. These expressions  break the permutation  
symmetry among
all
eigenvalues,  leaving
only a  permutation symmetry among the first-class eigenvalues   
together with
an
independent  permutation symmetry among the second-class  
eigenvalues.  The
coefficients
$a_i$ are nevertheless symmetric  in the whole set of eigenvalues  
and this is
precisely why  it is a more involved task to rewrite them in terms
of the corresponding supertraces.

It would seem that a reasonable starting point for the construction  
of  null
polynomials in the case of supermatrices  is the natural extension of
the definition  (\ref{POLCAR}) to  $ P(x) \rightarrow  Sdet (x{\cal I} -
M)$.
Nevertheless, this function  is not a polynomial but a ratio of  
polynomials, as
can be seen from
its
expression in terms of  the eigenvalues
\begin{equation}
 Sdet (x{\cal I} - M)= {(x-\lambda_1) \dots (x-\lambda_m) \over (x-{\bar
\lambda}_1) \dots
 (x-{\bar \lambda}_n) } .
\label{SDETEI}
\end{equation}

As  a preliminary step towards  the definition of the null polynomial
associated to
the supermatrix (\ref{SM}),  let us
introduce the standard null polynomials corresponding to the even
block-matrices
$A$ and $D$
\begin{equation}
a(x)= det ( x {\cal I}_m - A), \quad d(x)= det ( x {\cal I}_n - D),
\label{UCARPOL}
\end{equation}
respectively. 
Next, let us consider the general expression for
\begin{equation}
h(x):= Sdet (x{\cal I}_{m+n} - M),
\label{GCF0}
\end{equation}
which will be called the characteristic function in the sequel.  
From the two
alternative expressions
to calculate the superdeterminant, given in Eq.  (\ref{SDETM}),  we  
are able to
write the characteristic
function as
\begin{equation}
h(x) = {\tilde F(x)\over \tilde G(x)} = {F(x) \over G (x)},
\label{GCF}
\end{equation}
where the basic polynomials $\tilde F$, $\tilde G$, $F$ and $G$ are  
given by
\begin{eqnarray}
\tilde F(x) &=& det (d(x) (xI-A) - B adj(xI-D) C), \ \ {\tilde G}(x) =
(d(x))^{m+1}, \nonumber \\
F(x) &=& (a(x))^{n+1}, \ \ G(x) = det (a(x) (xI-D) - C adj (xI-A)B).
\label{BAPOL}
\end{eqnarray}
The above expressions are readily  obtained from Eqs.  
(\ref{SDETM}),  using
the
relation $(xI-F)^{-1} = [det(xI-F)]^{-1} adj
(xI-F)$ valid for any even matrix $F$. Notice that $\tilde F$ is
expressed in terms of the determinant of a $m \times m$ even
matrix, while $G(x)$ is the determinant of a $n \times n$ even
matrix.

In order to motivate the basic idea of our definition for the
characteristic polynomial of a supermatrix, let us consider the
simple case of a block-diagonal supermatrix $M \ (i.e. \ B = 0, C =
0)$. Here $h(x) = a(x)/d(x)$ and clearly the characteristic
polynomial is $P(x) = a(x) d(x)$, which is the product of the  
numerator and
the
denominator of the corresponding  superdeterminant. In fact we have
\begin{equation}
P(M) = \left( \matrix{ a(A) & 0\cr
0 & a(D) \cr}\right) \
\left(\matrix{ d(A) & 0\cr
0 & d(D)\cr} \right) \equiv 0, 
\label{EXCARP}
\end{equation}
because $a(A) = 0, d(D) = 0$. An analogous statement is obtained for a
supermatrix
written  in terms of its eigenvalues.
In the general case,  where  $h(x)$ is given by Eq.(\ref{GCF}), the  
numerator
of
the superdeterminant is  $\tilde F$  \  ($F$)  while the  
denominator is  \  $
\tilde G$  ($G$). These
observations lead to  the following definition of the characteristic
polynomial
of a supermatrix
\begin{equation}
{\cal P}(x) := \tilde F (x) G(x) = F(x) \tilde G(x),
\label{CPSM}
\end{equation}
in terms of  the basic polynomials $\tilde F, \tilde G, F$ and $G$,
 given in Eqs.(\ref{BAPOL}). For notational simplicity we will not  
necessarily
write explicitly the $x$-dependence on many of the polynomials  
considered in
the sequel.

Let us consider again  the block-diagonal case, this time  when  
$a(x)$ and
$d(x)$ have a common factor $f(x)$, i.e.
\begin{equation}
a(x) = f(x) a_1(x) , \ d(x) = f(x) d_1 (x).
\label{COMFAC}
\end{equation}
In this example, a null  polynomial is given by $P(x) = f(x) a_1  
(x) d_1(x),$
which is a polynomial of degree lower  than the product $a(x)
d(x)$.
Motivated by this fact,  together with the work of Ref. \cite{JAP2}, we
realize that there are some cases in which we can construct null
polynomials of lower degree than ${\cal P}(x)$, according to the  
factorization
properties of the basic polynomials $\tilde F, \tilde G, F, G$.

At this point it is important to observe that we do not have a
unique factorization theorem for polynomials defined over a
Grassmann algebra. This can be seen,  for example,  from the  
identity $x^2 =
(x+ z\alpha)(x- z\alpha)$, where $\alpha$ is an even Grassmann with
$\alpha^2 = 0$ and $ z$ is an arbitrary complex number.

The construction of the null polynomials of lower degree starts
from finding the divisors  of the pairs $\tilde
F, \tilde G, (F,G)$ which we denote by $R, (S)$ respectively. This
means that one is able to write
\begin{eqnarray}
\tilde F &=& R \tilde f, \ \  \tilde G = R \tilde g, \nonumber \\
F &=& Sf, \ \  G = Sg,
\label{FACTF}
\end{eqnarray}
where all polynomials are monic and also $\tilde f, \tilde g,f,g$  
are of lower
degree that their parents $\tilde F, \tilde G, F,G,  $  by  
construction. They
must satisfy
\begin{equation}
\tilde f/\tilde g = f/g,
\label{REDRAT}
\end{equation}
because of Eq. (\ref{GCF}).   The expressions in (\ref{FACTF})  
might not  be
unique.

Let us emphasize  that in the case of polynomials over
the complex numbers, when $R$ and $S$ are of maximum degree,   Eq.
(\ref{REDRAT}) would imply at most $\tilde f =
\lambda f, \tilde g = \lambda g$ with $\lambda$ being a constant.
Since we are considering polynomials over a Grassmann algebra,
this is not necessarily true as can be seen again in the above
mentioned identity $x/(x- z \alpha) = (x+ z \alpha)/x$, which we have
rewritten in a convenient way.

The above discussion leads  to the following
definition:  given an arbitrary $(m+n) \times (m+n)$ supermatrix  
$M$, with a
characteristic function $h(x)$  such that the polynomials  $\tilde  
F, \tilde G$
 have a common
factor   $R$  and the polynomials $ F, G$ have a common factor $S$,  
satisfying
Eqs.
(\ref{FACTF}) and (\ref{REDRAT}), then  a null polynomial of $M$   
is given by
\begin{equation}
P(x) := \tilde f(x) g (x) = f(x) \tilde g (x).
\label{NP}
\end{equation}
The  polynomial (\ref{NP}) is clearly of lower degree than ${\cal  
P}(x)$,
which is  just a particular case of the null polynomials (\ref{NP})  
 when
$R=S=1$.  We will
concentrate mostly on  (\ref{NP}) in the sequel.

\subsection*{6.2. Proof of the Cayley-Hamilton theorem for  
supermatrices}

In this section we show that the polynomial defined  in  
Eq.(\ref{NP}) does  in
fact
annihilates the supermatrix $M$. To this end, we first extend a lemma
often used to prove the 
Cayley-Hamilton theorem  for
ordinary matrices  \cite{DCHT}.  We briefly recall such lemma and
emphasize that it is independent of the matrix considered being a
standard matrix or a supermatrix. It goes as follows:
let $M$, $(xI-M)$ and $N(x)$ be $(m+n)\times (m+n)$
supermatrices,  where $M$ is independent of $x$. Let  $N(x)$
be a polynomial supermatrix  of degree $(p-1)$ in $x$, i.e.   $N(x) =
N_0x^{p-1}
+ N_1 x^{p-2} + .. + N_{p-1} x^0$, (where each $N_k,  \  k = 0,
\cdots , p-1, $ is a $(m+n)\times(m+n)$ supermatrix independent of  
$x$),  such
that
\begin{equation}
(x{\cal I}_{m+n}-M) N(x) = P(x){\cal  I}_{m+n},
\label{NSM}
\end{equation}
where $P(x) = a_0 x^p + a_1 x^{p-1} + \cdots + a_n x^0$ is a
numerical polynomial of degree $p$. Then, one can prove that  $P(M)  
:= a_0 M^p
+
a_1 M^{p -1} + \cdots + a_n {\cal I}_{m+n} \equiv 0$.
The proof follows by
comparing the independent powers of $x$ in Eq. (\ref{NSM}) and then  
explicitly
computing $P(M)$ \cite{DCHT}.

In the standard case,  the polynomial matrix
$N(x)$ is just given by $N(x) = adj (xI-M) = det(xI-M)
(xI-M)^{-1}$, and $P(x) = det (xI-M)$. In the case of a supermatrix  
we do not
have an obvious
generalization either of the matrix $adj (xI-M)$ or
of $det (xI-M)$. Nevertheless, following the analogy as close as
possible we define
\begin{equation}
N(x) := P(x) (xI-M)^{-1},
\label{DEFN}
\end{equation}
where $P(x)$ is the polynomial
introduced in Eq.(\ref{NP}) of the previous section.

 The challenge now is to prove
that $N(x)$, which trivially satisfies the Eq. (\ref{NSM}), is indeed a
polynomial matrix. In this way we would have proved that $P(M) =
0$ according to the property stated after Eq.(\ref{NSM}).

To begin with,  we
show that
the blocks corresponding to the inverse supermatrix $ (x{\cal I}_{m+n}-
M)^{-1}$ can
be written in a compact form as
\begin{eqnarray}
(x {\cal I}_{m+n}-M)^{-1}_{ij} &=& - {1\over \tilde F}{\partial  
\tilde F\over
\partial A_{ji}},\quad (x {\cal I}_{m+n}-M)^{-1}_{i \alpha } =   
{1\over G}
{\partial
G\over \partial C_{\alpha i}} \label{XIMMI}  \\
(x {\cal I}_{m+n}-M)^{-1}_{\alpha j} &=&  {1\over \tilde F}  
{\partial \tilde
F\over
\partial B_{j \alpha }}, \quad (x {\cal  
I}_{m+n}-M)^{-1}_{\alpha\beta} = -
{1\over G}
{\partial G\over \partial D_{\beta\alpha}},
\label{XIMMA}
\end{eqnarray}
where $A_{ij},B_{j  \alpha},C_{\alpha j} $and $D_{\alpha \beta}$
are the entries of the supermatrix  $M$ defined in Eq. (\ref {SM})
and  $\tilde F$, \  $G$,  are the polynomials given in Eqs.  
(\ref{BAPOL}). The
derivative with respect to an odd Grassmann number is taken to be a left
derivative,
defined such that  $\delta\tilde F
\equiv \delta B_{j\alpha} {\partial\tilde F\over \partial
B_{j\alpha}}$. The proof of the above equations begins with  the
calculation of  $(x {\cal I}_{m+n}-M)^{-1}$ in block form, with
the results
\begin{eqnarray}
(x {\cal I}_{m+n}-M)^{-1}_{11} &=& ((x {\cal I}_{m}-A)-B (x {\cal
I}_{n}-D)^{-1} C)^{-1},
\nonumber  \\
(x {\cal I}_{m+n}-M)^{-1}_{12} &=&-(x {\cal I}_{m}-A)^{-1}B ((x {\cal
I}_{n}-D)
- C(x {\cal I}_{m}-A)^{-1} B)^{-1},  \nonumber \\
(x {\cal I}_{m+n}-M)^{-1}_{21} &=& -(x{\cal I}_{n}-D)^{-1}C ((x {\cal
I}_{m}-A)-B (x {\cal I}_{n}-D)^{-1}C)^{-1}, \nonumber \\
(x {\cal I}_{m+n}-M)^{-1}_{22} &=& ((x {\cal I}_{n}-D)-C (x {\cal
I}_{m}-A)^{-1} B)^{-1}.
\label{INVXIMM}
\end{eqnarray}
Here, the subindices 11, 12, 21 and 22 denote the corresponding
$m\times m, m\times n, n\times m$, and $n\times n $ blocks. The  
above block
form in Eq. (\ref{INVXIMM})
has  the same structure  as in  the  classical case.  Let us concentrate
now in the $11$ block. Rewriting all the inverse matrices of  the first
Eq.(\ref{INVXIMM}) in
terms of their adjoints, together with the corresponding  
determinants,  we
obtain
\begin{equation}
(x {\cal I}_{m+n}-M)^{-1}_{11}= {d\over \tilde F} adj ( (x {\cal  
I}_{m}-A)d -
Badj
(x {\cal I}_{n}-D) C).
\label{INV11}
\end{equation}
On the other hand, using the basic property
\begin{equation}
\delta det Q = Tr (adj Q \delta Q),
\label{DELQ}
\end{equation}
valid for any even matrix $Q$, we can  calculate the change of  
$\tilde F$ with
respect to $A_{ij}$, keeping constant all other entries, obtaining
\begin{equation}
\delta\tilde F = - d \left[ adj ((x{\cal I}_{m}-A)d - B adj
(x {\cal I}_{n}-D)C\right]_{ij} \ \delta A_{ji},
\label{DELFT}
\end{equation}
which can be written as
\begin{equation}
{\partial\tilde F\over \partial A_{ji}} = - d
\left[ adj ((x {\cal I}_{m}-A)d - B adj (x {\cal I}_{n}-D)  
C)\right]_{ij}.
\label{DTFA}
\end{equation}
The comparison of Eq. (\ref{DTFA}) with Eq. (\ref{INV11}) completes  
the proof
of
the first
relation in Eq. (\ref{XIMMI}). The  proof for the remaining Eqs.
(\ref{XIMMI}-\ref
{XIMMA}) is carried along similar steps.

We observe  that the conditions for the existence of
$(x {\cal I}_{m+n}-M)^{-1}$ are the  same as those for the  
existence of $Sdet
(x {\cal I}_{m+n}-M)$ which read $det(x {\cal I}_{m}-A) \not= 0$  
and $det(x
{\cal I}_{n}-D) \not=
0$. Since $x$ is a generic even Grassmann variable, we will assume that
this is always the case. In this way, the term
$((x {\cal I}_{m}-A) - B(x {\cal I}_{n}-D)^{-1} C)^{-1}$, for  
example, can
always be calculated as
$( {\cal I}_{m}-(x {\cal I}_{m}-A)^{-1}B (x {\cal I}_{n}-D^{-1})  
C)^{-1} (x
{\cal I}_{m}-A)^{-1}$. The factor on the
left can be thought as a series expansion of the form $1/(1-z) = 1  
+z+z^2+
\cdots ,$ with $z= (x {\cal I}_{m}-A)^{-1}B
(x {\cal I}_{n}-D)^{-1} C$. Moreover, the series will stop at some power
because $z$ is a nilpotent  matrix.

Now we
are in position  to show  the principal result of this section  
which is that
$N(x)
:= P(x)
 (x {\cal I}_{m+n}-M)^{-1}$, with $P(x)$ given in
Eq.(\ref{NP}),  is a polynomial supermatrix.
Let us consider the block-element $11$  of
$N(x)$ to begin with. According to Eqs. (\ref{XIMMI}-\ref{XIMMA})
 together with Eq. (\ref{FACTF}), this block can be written as
\begin{equation}
N_{ij} = - g {\partial \tilde f\over \partial A_{ji}} -
{g \tilde f \over R} {\partial R\over \partial A_{ji}}.
\label{N11}
\end{equation}
The first term of the RHS in Eq.(\ref{N11}) is clearly of  
polynomial character.
In
order to see that  the second term is also polynomial,  we make use  
of the
property
\begin{equation}
{\partial ln \tilde G\over \partial A_{ji}} = 0 = {\partial ln
R\over \partial A_{ji}} + {\partial ln \tilde g \over \partial
A_{ji}},
\label{DTGDA}
\end{equation}
which follows from the factorization $\tilde G = R \tilde g$,
together with the fact that $\tilde G$ is just a function of
$D_{\alpha\beta}$, according to the first  Eq. (\ref{BAPOL}). In  
this way, and
using  also the
Eq.(\ref{REDRAT}),  we obtain
\begin{equation}
N_{ij} = f {\partial \tilde g\over \partial A_{ji}} -
g {\partial \tilde f\over \partial A_{ji}},
\label{N11P}
\end{equation}
which leads to the conclusion that the block-matrix $N_{ij}$ is
indeed polynomial. The proof for $N_{\alpha i}$ runs along the
same lines, except that now the derivatives are taken with
respect to $B_{i\alpha}$ and that we have to use ${\partial ln
\tilde G\over \partial B_{i\alpha}}= 0$, instead of Eq. (\ref{DTGDA}).
The remaining terms $N_{i\alpha}$ and $N_{\alpha\beta}$ can be
dealt with in analogous manner, by considering the derivatives of
$ G =S g$ with respect to $C_{\alpha i}$ and
$D_{\beta\alpha}$, and by replacing the condition (\ref{DTGDA}) by
${\partial ln F\over \partial C_{\alpha i}} = 0 $ and ${\partial
ln F\over \partial D_{\beta\alpha}} = 0 $ respectively. The
results are again of the form (\ref{N11P}), the only difference been the
variables with respect to which the derivatives are taken.

Finally,  we can state the
following extension of the Cayley-Hamilton theorem to the case  of
supermatrices \cite{URRU3},  \cite{URRU4},  \cite{URRU5}.

{\bf Theorem } (Extended Cayley-Hamilton Theorem)  Let $M$ and  
$(xI-M)$ be
$(m+n)\times (m+n)$ supermatrices, with  $x $ being a generic even  
Grassmann
variable. Let also $Sdet(x{\cal I}_{m+n}-M)=\tilde
F/\tilde G=F/G$,  where the polynomials $\tilde F, \tilde G, F$ and  
$G$ are
given in Eqs.(\ref{BAPOL}). Then, for any  common factor $R$ such  
that $\tilde
F = R
\tilde f, \  \tilde G = R \tilde g$ and $S$ such that $F = Sf, \  G  
= Sg$,
where $\tilde f/\tilde g = f/g,  $ the polynomial $P(x)=\tilde f(x)  
g (x) =
f(x) \tilde g (x)$  annihilates $M$, i.e.  $P(M)=0.$

\subsection*{6.3.  Examples of null polynomials for supermatrices}

Here we  present two simple examples of null polynomials, 
constructed  according to the procedure stated in the last section

\paragraph*{ The case of $(1+1)\times(1+1)$ supermatrices.}

Let us consider the supermatrix (\ref{SM22}) with ${\bar d} \neq 
{\bar q}$. Here the bar denotes the complex component of an even
Grassmann number, called the body in the literature. From Eqs.
(\ref{BAPOL}) we obtain the following basic polynomials
\begin{eqnarray}
{\tilde F}&=& (x-q)(x-p) -\eta \theta, \quad {\tilde G}= (x-q)^2,  
\nonumber
\\
{F}&=& (x-p)^2,  \quad {G}= (x-q)(x-p)+ \eta \theta,
\label{BP22}
\end{eqnarray}
The above functions can be  rewritten as follows
\begin{eqnarray}
{\tilde F}&=& \left( x-p + {\eta \theta \over q-p} \right) 
\left(  x-q - {\eta \theta \over q-p}  \right), \nonumber \\
{\tilde G}&=& \left( x-q + {\eta \theta \over q-p} \right) 
\left(  x-q - {\eta \theta \over q-p}  \right), \nonumber \\
F &=& \left( x-p + {\eta \theta \over q-p} \right) 
\left(  x-p - {\eta \theta \over q-p}  \right), \nonumber \\
{G}&=& \left( x-q + {\eta \theta \over q-p} \right) 
\left(  x-p - {\eta \theta \over q-p}  \right).
\label{FACT22}
\end{eqnarray}
Thus, the factorization properties of Eq.(\ref{FACTF})
are realized  with
\begin{eqnarray}
R &=& \left(  x-q - {\eta \theta \over q-p}  \right), \quad 
S = \left(  x-p - {\eta \theta \over q-p}  \right), \nonumber \\ 
{\tilde f}= f &=& \left( x-p + {\eta \theta \over q-p} \right), 
\quad {\tilde g}= g =\left( x-q + {\eta \theta \over q-p} \right). 
\label{FACT221}
\end{eqnarray}
In this way, using the definition (\ref{NP}), we recover  the null  
polynomial 
of minimum degree given in Eq.(\ref{PCSM2}).

\paragraph*{ The case of $Osp(1|2; {\cal C})$ supermatrices.}
Another simple example corresponds to the case of supermatrices
belonging to the supergroup $Osp(1|2; {\cal C})$, which are relevant
in the description of de Sitter supergravity in $2+1$ dimensions 
\cite{URRU1}. We consider a $(2+1) \times (2+1)$ realization of this
supergroup  defined by the set of all supermatrices $M$ which 
leave invariant the supersymplectic form $H$ 
\begin{equation}
M^{\cal T} H M=H, \quad 
H=\pmatrix { \ 0 & 1 & 0 \cr -1 & 0 & 0  \cr \ 0 & 0 & 1 },
\label{OSP}
\end{equation}
where ${\cal T}$ denotes the supertransposed. The  supermatrices in
Eq.(\ref{OSP}) 
can be parametrized in the following way
\begin{equation}
M=\pmatrix {  A  &  \xi  \cr \chi^T  & a  }, \quad 
\xi= \pmatrix { x_1  \cr  x_2 }, 
\label{PARAM}
\end{equation}
where $A$ is a $(2\times 2)$ even matrix;  $x_1, x_2$  are arbitrary
odd Grassmann numbers and the superindex $T$ denotes standard
transposition. The condition (\ref{OSP}) translates into  the following
relations among the elements of Eq.  (\ref{PARAM})
\begin{equation}
\chi^T=\xi^T E A, \quad a=1+x_1 x_2, \quad det(A)= 1- x_1 x_2,
\label{CONST}
\end{equation}
where $E$ denotes the $2\times 2$ antisymmetric block of $H$ in 
Eq.(\ref{OSP}). Using the algorithm of the previous section, we
conclude  that the  unique irreducible expression for the 
characteristic function is
\begin{equation}
h(x)= { x^2 -(1 + Str(M))x +1) \over {x -1 }},
\label{HOSP}
\end{equation}
which  means $f = {\tilde f}, g = {\tilde g}$. The  null polynomial
of minimum degree is then \cite{URRU1} 
\begin{equation}
P(x)= f g = x^3 -( 2 + Str(M))(x^2-x) -1.
\label{MPOSP}
\end{equation}

\subsection*{6.4.  Two examples of   Mandelstam identities for  
supermatrices}

 As a first  step towards the  search  of  an algorithm to produce   
the MI for
supermatrices, one may try to directly extend the procedure
developed in section  5.3 for the case of ordinary matrices. The  
starting point
now
will be the null polynomials constructed in section 6.1., which can be
rewritten in terms of a finite number of supertraces.  In relation  
to this,  we
still suffer a main drawback which
is the lack of
knowledge of a recurrence that  would allow  to obtain a closed  
expression for
the
coefficients of the null polynomial in terms of supertraces, in a manner
similar to the
standard case.  Another  new feature is that the null polynomials are not
monic any more,
in such a way that $a_0$ would now be a function of the supertraces. This
property
will effectively raise the degree of homogeneity of the  
supermatrices in  the
corresponding
CHI. Nevertheless, such identities will be homogeneous of some  
degree, say $t$
for
example. This will allow us to make the following definition of  the
corresponding generic MI
\begin{equation}
Str \left( P_{M_1+M_2+ \dots + M_t}(M_1+M_2+ \dots + M_t)|_{red} M_{t+1}
\right)=0.
\label{MISMF}
\end{equation}
 \paragraph*{ The case of $(1+1)\times(1+1)$ supermatrices.} In  
this situation,
the
null polynomial (\ref{PCSM2})  can be rewritten as
\begin{equation}
Str(M) M^2 - (Str(M^2)) M + {1 \over 3}\left( Str(M^3) - 
{Str(M)}^3 \right){\cal I}_{2+1}= 0,
\label{CHSM}
\end{equation}
in terms of supertraces, with  $t=3$. Using the definition  
(\ref{MISMF}) we
obtain
\begin{eqnarray}
&& Str(A) \left( Str(BCD) + Str( CBD) \right)
+ Str(B) \left( Str(ACD) + Str( CAD)\right) \nonumber \\
&& Str(C) \left( Str(ABD) + Str( BAD)\right)
+ Str(D) \left( Str(ABC) + Str( BCA)\right)\nonumber \\
&& -2 Str(AB) Str(CD)-2 Str(BC) Str(AD)-2 Str(AC) Str(BD)\nonumber \\ 
&&-2 Str(A) Str(B) Str(C) Str(D) =0,
\label{MISM11}
\end{eqnarray}
which corresponds to a symmetric MI of order four. We have verified this
identity using
Mathematica.

\paragraph*{ The case of $Osp(1|2; {\cal C})$ supermatrices.} This is an
example of a restricted
MI, which is the supersymmetric analogue of the identity  
(\ref{MISU}), valid
for the group $SU(2)$. Starting from the null polynomial   
(\ref{MPOSP}) for the
supermatrix $M_1$, multiplying this equation
by $M_1^{-1} M_2$ and taking the supertrace we are left with
\begin{equation}
Str (M_2 M_1^2) -( 2+ Str(M_1)) ( Str (M_1 M_2) - Str(M_2)) - Str ( M_2
M_1^{-1})=0.
\label{SUSYMI2}
\end{equation}
The above identity  has been useful in  the identification of the  
true degrees
of freedom
on  one sector of  $2+1$ super de Sitter gravity.  For one genus of  
the generic
spatial surface,
the most general Wilson loop variables are the infinite set of  
supertraces: $
Str ( M_1^{p_1} M_2^{q_1} M_1^{p_2} M_2^{q_2} \dots M_1^{p_n}
M_2^{q_n}
\dots ) $, for any  integer $p_i, \ q_i$. Using  the MI  
(\ref{SUSYMI2}),  it is
possible to reduce the above infinite set of observables to only  
five complex
quantities, which are  $Str(M_1), Str(M_2),
Str( M_1 M_2), Str ( M_1 M_2^2)$ and $ Str(M_1M_2 M_1^2M_2^2)$  
\cite{URRU1}, in
complete analogy with $2+1$ de Sitter gravity \cite{NEREZE}.

\section*{7.  Summary  and open problems}

We have presented a rather sketchy review of the loop space  
formulation of
gauge theories,
which does not make full justice  to all   the numerous achievements and
applications   that this method has produced so far. We have tried to
incorporate a  reasonable, but certainly not exhaustive,  list of  
references
which remedy
this situation, offering the reader a detailed version in each situation.

Our major emphasis has been in the discussion of the formulation of the
Mandelstam identities, which appear as
unavoidable constraints either  among  the Wilson loop variables or  
among  the
quantum loop states,  that constitute  the natural degrees of  
freedom of the
method in the classical and quantum situation, respectively. In the  
case of
pure gauge theories over a Lie algebra, we have shown the  
equivalence between
the generic Mandelstam identities, for a given dimension of the
representation, and the identities arising from the application of the
Cayley-Hamilton theorem
to the matrices of  such  representation. We have provided an  
algorithm to go
in either
direction.  The main thrust of this development has been it extension to
provide  a general procedure to construct the generic Mandelstam  
identities in
the case of  pure gauge theories defined over a super Lie algebra,  
as a first
step to deal with fully supersymmetric theories. A
previous step in this direction has been  the formulation  and  
proof of the
Cayley-Hamilton theorem in the case of supermatrices.
The extension of this  theorem proceeded  in two steps: (i) the  
identification
and definition of a  characteristic polynomial for supermatrices  
and (ii) the
proof that each supermatrix annihilates the polynomial previously  
defined.
Furthermore, starting from the characteristic function  
(\ref{GCF0}), we have
described  a systematic method  for constructing null polynomials for
supermatrices.  The
resulting Cayley-Hamilton identities can be subsequently  used to  
derive the
corresponding Mandelstam identities. Two simple examples were  
presented. The
construction of  the
Mandelstam identities for arbitrary restricted groups or  
supergroups remains
still an open problem.

Another very interesting question in the loop space formulation of gauge
theories is the inclusion of fermions, which  has been the subject  
of  recent
investigations. There are at least  three ways of  approaching this  
problem:
(i)  one is to consider the fermions as standard matter to be  
coupled to the
gauge theory, (ii) other possibility is
to consider them as pieces of a superconnection, as has been done  
in these work
and (iii) the final possibility is to introduce the
fermions  in a fully supersymmetric theory, as partners of integer  
spin fields.
In the first case, several matter fields like electrons  
\cite{GAMBFORT} and
quarks \cite{GAMBSET} have been taken onto account  in the loop  
space picture.
Also, the introduction of fermions in Einstein gravity has been  
considered in
Ref. \cite{MORROV}. The basic idea in these works is to define   
additional
gauge invariant variables, besides the Wilson loops, represented by  
open paths
which start and end up in the fermions.  As mentioned previously in  
the text,
the possibility (ii) has been  already considered   in the case
of $2+1$ super Chern-Simons theories and also in  the recent  
discussion of
$3+1$ canonical
supergravity in the loop space approach.  From a general point of  
view, there
is still lacking
a proof of the equivalence between the proposed loop space  
representation of
these theories and the standard connection-matter formulation of  
them. This
will require, among many other
developments, the construction of
generalized Mandelstam identities including the open path variables  
together
with the generalization of Giles theorem to the superconnection  
case. Finally,
the alternative (iii)
has recently being explored in Ref. \cite{AWMAN}, where the fully
supersymmetric Wilson
loop has been constructed in terms of chiral superfields and  
supercurrents in
superspace.

\

The author acknowledges the organizers of  the  Fifth Workshop on  
Particles and
Fields, in particular J.C. D'Olivo, M.A. P\'erez and C. Ram\'\i rez, for
inviting him to present the lectures
in which this  contribution is based.  Support from the grants
UNAM-DGAPA-100694 and  CONACyT  3544-E9311  is gratefully acknowledged.

\end{document}